\documentclass[sigconf]{acmart} 
\AtBeginDocument{%
  \providecommand\BibTeX{{%
    \normalfont B\kern-0.5em{\scshape i\kern-0.25em b}\kern-0.8em\TeX}}}

\copyrightyear{2023} 
\acmYear{2023} 
\setcopyright{acmlicensed}\acmConference[CHI '23]{Proceedings of the 2023 CHI Conference on Human Factors in Computing Systems}{April 23--28, 2023}{Hamburg, Germany}
\acmBooktitle{Proceedings of the 2023 CHI Conference on Human Factors in Computing Systems (CHI '23), April 23--28, 2023, Hamburg, Germany}
\acmPrice{15.00}
\acmDOI{10.1145/3544548.3581224}
\acmISBN{978-1-4503-9421-5/23/04}

\usepackage{listings} 

\usepackage{pdfpages}
\usepackage{subcaption}

\usepackage{xcolor}

\newcommand{\eg}{{e.g.,}}
\newcommand{\etal}{{et~al.}}

\begin{document}

\title[Enabling People Who Stutter to Better Use Speech Recognition]{From User Perceptions to Technical Improvement: Enabling People Who Stutter to Better Use Speech Recognition}



\author{Colin Lea}
\email{colin_lea@apple.com}
\affiliation{%
  \institution{Apple}
  \city{Pittsburgh, Pennsylvania}
  \country{USA}}

\author{Zifang Huang}
\email{zhuang7@apple.com}
\affiliation{%
  \institution{Apple}
  \city{Pittsburgh, Pennsylvania}
  \country{USA}}

\author{Jaya Narain}
  \email{jnarain@apple.com}
\affiliation{%
  \institution{Apple}
  \city{Cupertino, California}
  \country{USA}}

\author{Lauren Tooley}
\email{ltooley@apple.com}
\affiliation{%
  \institution{Apple}
  \city{Baltimore, Maryland}
  \country{USA}}

\author{Dianna Yee}
\email{dianna_yee@apple.com}
\affiliation{%
  \institution{Apple}
  \city{Seattle, Washington}
  \country{USA}}

\author{Tien Dung Tran}
\email{dung_tran@apple.com}
\affiliation{%
  \institution{Apple}
  \city{Cupertino, California}
  \country{USA}}

\author{Panayiotis Georgiou}
\email{panayiotis_georgiou@apple.com}
\affiliation{%
  \institution{Apple}
  \city{Cupertino, California}
  \country{USA}}

\author{Jeffrey P. Bigham}
\email{jbigham@apple.org}
\affiliation{%
  \institution{Apple}
  \city{Pittsburgh, Pennsylvania}
  \country{USA}}

\author{Leah Findlater }
\email{lfindlater@apple.com}
\affiliation{%
  \institution{Apple}
  \city{Seattle, Washington}
  \country{USA}}

\renewcommand{\shortauthors}{Lea et al.}

\begin{abstract}
Consumer speech recognition systems do not work as well for many people with speech differences, such as stuttering, relative to the rest of the general population. However, what is not clear is the degree to which these systems do not work, how they can be improved, or how much people want to use them. In this paper, we first address these questions using results from a 61-person survey from people who stutter and find participants want to use speech recognition but are frequently cut off, misunderstood, or speech predictions do not represent intent. In a second study, where 91 people who stutter recorded voice assistant commands and dictation, we quantify how dysfluencies impede performance in a consumer-grade speech recognition system. Through three technical investigations, we demonstrate how many common errors can be prevented, resulting in a system that cuts utterances off 79.1\% less often and improves word error rate from 25.4\% to 9.9\%. 
\end{abstract}




\maketitle

\section{Introduction}
Performance of speech recognition systems has improved substantially in recent years, leading to widespread adoption of speech input across computing platforms. Speech is especially important for devices with limited or no screen real estate, such as smart speakers (e.g., Amazon Echo, Google Nest Audio)
and hearables (e.g., Pixel Buds, Apple AirPods), where it is used 
for everyday tasks like playing music, searching for facts, and controlling smart devices in the home~\cite{AmmariCHI21}.
Yet, speech interaction presents accessibility barriers for many people with communication disabilities such as stuttering, dysarthria, or aphasia~\cite{projecteuphonia,ballati2018assessing,fok2018towards,roper2019speech}.
In this work, we investigate user experiences and speech recognition system performance with one large population in this demographic of communication disability: people who stutter (PWS).

Stuttering, also called stammering~\cite{ward2011cluttering}, impacts approximately 1\% of the world's population, although estimated incidence ranges from 2\% to 5\% for sub-populations such as children and males~\cite{craig2002epidemiology}.  
Stuttering is a breakdown in speech fluency that can affect the rate and flow of speech and can include dysfluency types such as sound repetitions (``p-p-pop''), syllable repetitions (``be-be-become''), word or phrase repetitions (``mom, mom, mom''), sound prolongations (``mmmmmom''), and audible or silent blocks (pauses or breaks in speech) \cite{sander1963frequency, SSI}. 
Frequent interjections (``um'', ``eh'') are also common. 
The rate, duration, and distribution of these dysfluencies varies substantially between people and across contexts~\cite{tichenor2021variability}. 

Research on speech technology for PWS has largely focused on technical improvements to automatic speech recognition (ASR) models~\cite{shonibare2022enhancing,Mendelev2020,gondala2021error,mitra2021,HeemanInterSpeech16}, dysfluency detection~\cite{lea2021sep,kourkounakis2020detecting,Mahesha2016,DasNSRE2020}, and dataset development~\cite{FluencyBank,UCLASS,lea2021sep,KSoF}. 
This body of work has largely lacked a human-centered approach to understanding the experiences that PWS have with speech recognition systems~\cite{clark2020speech}, which could in turn inform how to prioritize and advance technical improvements. A recent exception from Bleakley et al.~\cite{bleakley2022exploring} conducted an interview and diary study on the use of voice assistants with 11 PWS. That study provides an initial understanding of user experiences with a voice assistant, identifying issues with the device timing out and that social pressure can influence use. There are still open questions, however, in how PWS more broadly experience speech technologies (i.e., not just voice assistants), the prevalence of accessibility challenges encountered, how these experiences relate to stuttering severity and specific dysfluencies, and how speech recognition systems perform---and can be improved---in the context of these reported challenges.


In this paper, we report on two studies designed to address these questions of user experience and system performance for both voice assistants (VAs) and dictation: a survey on the use of speech technology with 61 PWS, and an investigation of VA and dictation system performance on speech data collected from a larger set of 91 PWS (53 of whom overlapped with the survey). 
Both studies characterize what attributes, including prevalence of dysfluency types, tend to impact experience and cause recognition errors. Furthermore each study highlights ways in which speech systems could be improved to support the nuanced speech patterns of PWS.
Overall, the findings show that speech technologies work reasonably well for some PWS with mild speech dysfluency patterns, as reflected in both the survey and performance data of those. However, people with more moderate or severe patterns encountered high truncation (e.g., > 20\%) and word error rates (13.6\% to 49.2\%)---which is also reflected by many survey participants. Motivated by these results, we describe and evaluate three technical solutions (two new and one previously published~\cite{mitra2021}) that apply production-oriented improvements to a consumer-grade ASR system. These solutions reduce truncation rates by 79.1\% and improve word error rates in transcribed speech from 25.4\% to 9.9\% for a set of participants with moderate to severe dysfluent speech.

Overall, this paper contributes an understanding of the subjective experiences that PWS have with VAs and dictation across a range of stuttering severities, confirms and reflects those experiences through performance evaluation with a commercial ASR system, and demonstrates the utility of three technical improvements that can be applied with relatively small adaptations to existing systems.
While continued effort is needed to improve these technologies for people with speech disabilities, as frequently noted in popular press (\eg ~\cite{Slate,USAToday,WSJ}), our solutions start to narrow the performance gap for many PWS and potentially improve performance for people with other speech disabilities as well.

\section{Background \& Related Work}

\subsection{Speech Recognition Systems}
\label{sec:rw-speechrecsystems}

In this section, we describe common pipelines for Voice Assistants and Dictation Systems, to build an understanding of how people use them and how the underlying models work. 
In Section~\ref{sec:survey_results} we describe how PWS use and perceive these systems and in Section~\ref{sec:asr_results} we demonstrate how dysfluencies impact performance of constituent models.

\begin{figure*}[t!]
    \centering
    \includegraphics[width=0.8 \paperwidth]{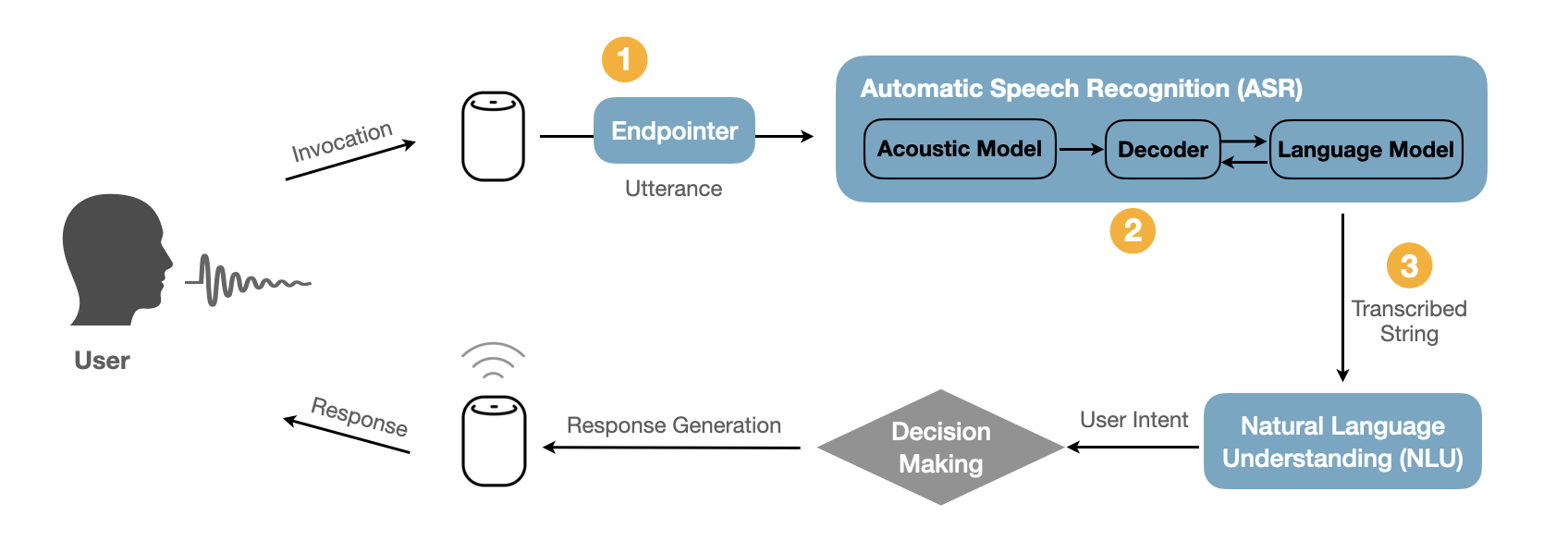}
    \caption{Simplified workflow and components of a voice assistant. Figure adapted from \cite{ranantha2020intent}. 
    In this work we seek to understand performance on speech from people who stutter.
    In Section~\ref{sec:asr_results}  we report on baseline performance of the endpointer and ASR models for dysfluency speech, and investigate three interventions to better support PWS at different points in the workflow: (1) endpointer tuning; (2) ASR decoder tuning; and (3) refinement of dysfluencies in the transcribed ASR output.
    }
    \label{fig:VA Illustration}
    \Description{A pipeline with each element of a voice assistant. A user speaks into a device to invoke the VA, an endpointer model monitors that speech and decides when to clip the audio into an utterance. 
    The ASR model is run on that utterance and outputs a text string, the NLU model takes the transcribed string and outputs the user intent, and a decision making model determines how the intent translates to a system response which is displayed to the user. }
\end{figure*}

\subsubsection{Overview of Speech Recognition Systems}

\paragraph{Voice Assistants (e.g., Amazon Alexa, Apple Siri, and Google Assistant)} These systems listen for spoken questions or commands. For example, ``What is the weather?'' or ``Set a timer for five minutes.'' 
A typical user flow is as follows, where a simplified view is shown in Figure~\ref{fig:VA Illustration}. 
A user may initiate a query by saying a \textit{wake word}, such as ``Alexa'', ``Hey Siri'', or ``OK Google.'' This may also be called \textit{invocation}. 
They then start vocalizing their command. 
As the user speaks, an \textit{ASR} model starts to transcribe their words and simultaneously an \textit{endpointer} model---also called end-of-speech detector---detects whether the user has finished speaking.\footnote{For simplicity, Figure \ref{fig:VA Illustration} shows the endpointer and ASR in sequence. In practice, endpointers range from being independent of an ASR \cite{Maas2018DNNEndpointer} to fully integrated \cite{li2020RNNTEndpointer}.}
A \textit{Natural Language Understanding (NLU)} model then takes the transcribed phrase and any metadata from the ASR and identifies the user's intent, which is then sent to the \textit{Decision Making} component to perform the appropriate action. 
Based on the user's query, the VA may respond with information, perform a command, ask the user a follow-up question (i.e., ``is the alarm request for AM or PM''), or notify the user that the query could not be understood. 



There are alternative ways of interacting with a VA. Instead of using a speech-based invocation mechanism on a phone or tablet, a user may hold down a button on the side of the device (``Hold to Speak'' on iOS) or squeeze the side of the phone (some Android devices). 
VAs can also be invoked using accessibility features such as Accessibility Menu on Android or Assistive Touch on iOS.
Instead of speaking, a user can type a command with a keyboard using ``Type with Alexa'' or ``Type to Siri.'' This bypasses the ASR component and may directly process the input text through NLU. 

\paragraph{Dictation Systems} These systems provide an alternative to using a keyboard by converting a user's speech to text for tasks such as composing text messages, notes, or emails. 
A common use case is to verbalize a text while in a car. 
Typically, a user starts dictation by manually clicking a button on a device, instead of via a wake word, and they finish by manually clicking a button again, instead of relying on an endpointer.   
An ASR model may transcribe their speech in real-time and visualize the text if the device has a screen. 
A \textit{punctuation} model may add grammatical structure to spoken words, like adding commas and periods.
Finally, if the ASR model yields low confidence in a predicted word, this word may be highlighted on screen along with suggestions for replacement. 
A user can then manually revise their statement or click on low confidence words for suggestions. 



\subsubsection{Model Descriptions \& Implications of Dysfluencies}

To refine speech technologies for PWS, we focus on endpointer and ASR models independently. Endpointer models are optimized to balance the trade-off between cutting someone off too early and having a long delay after their speech, and implicitly or explicitly rely on cues such as how much silence has elapsed since the last word (e.g., \cite{Maas2018DNNEndpointer,li2020RNNTEndpointer}). These models are largely optimized on data from people who do not stutter, and are tuned to respond to the user with as low latency as possible. 
For PWS, as described by Bleakley \etal~\cite{bleakley2022exploring} and in Sections \ref{sec:survey_results} and \ref{sec:asr_results}, endpointer models commonly cut off before the user is done speaking, especially during blocks where there may be a long pause or gasp, or even partial word repetitions where the vocalization may be indecipherable.

At a high level, ASR models (e.g., \cite{yi2021hybridASR, hinton2012HybridASR,chan2015ListenAttendSpell,graves2012RNNT}) commonly consist of an acoustic component that encodes low-level phoneme or phone-like patterns, a language component that acts as a prior over word sequences, in some cases a pronunciation component that describes the phonetic pronunciation of words (e.g., for hybrid ASRs \cite{hinton2012HybridASR, Sak2013HybridModelComponents, le2020hybridASR}),
and a decoder that may map these together and generate candidate transcriptions. 
For PWS, we hypothesize that sound, syllable, or partial word repetitions fed into the acoustic component may have lower confidence phone- or phone-like outputs, as these models may not be trained with large amounts of dysfluent speech.
Similarly, the language component may not be trained with word or phrase repetitions, meaning these repetitions may have lower likelihood and be misrecognized. 
Furthermore, interjections mid-word or partial word repetitions may cause errors when paired with a pronunciation dictionary and decoder, as there may be spurious phone-like units interspersed with known words. 

It is worth noting that while VAs and Dictation often rely on the same underlying ASR system, a VA's model may be tuned to work better for a smaller set of assistant commands~\cite{gondala2021error}, whereas Dictation may be biased towards free speech from arbitrary topics. 
In addition, for VAs, an NLU model (e.g., Chen \etal~\cite{chen2019active}, which is used in this paper) may take an erroneous transcription and map it to the correct action using context and error cues. 
This means that individual ASR errors may play a more important role in Dictation than for VA tasks.

\subsubsection{Dysfluent Speech Recognition}

Technical work on improving speech assistants for PWS has focused on ASR models~\cite{shonibare2022enhancing,Mendelev2020,gondala2021error,mitra2021,Tripathi2018,Tripathi2020,HeemanInterSpeech16}, stuttering detection~\cite{LeaStutterDetection2021}, dysfluency detection or classification~\cite{lea2021sep,kourkounakis2020detecting,Mahesha2016,DasNSRE2020,riad2020identification}, clinical assessment~\cite{BayerlTSD2020}, and dataset development~\cite{FluencyBank,UCLASS,lea2021sep,KSoF}.
Shonibare \etal ~\cite{shonibare2022enhancing} and Mendelev \etal  ~\cite{Mendelev2020} investigate training end-to-end RNN-T ASR models on speech from PWS. 
Shonibare \etal\ introduces a detect-then-pass approach that incorporates a dysfluency detector where audio frames with dysfluencies are ignored entirely by the RNN-T decoder. 
Examples in the paper demonstrate how this approach can be used to ignore some partial-word repetitions and blocks.
Alharbi \etal~ \cite{AlharbiWOCCI17,alharbi2018lightly} focused on stuttered speech from kids that incorporates the structure of repetitions and other dyfluencies into an augmented language model that is better at including dysfluencies in a transcription. 
In our work, we focus on solutions, like Mitra \etal's \cite{mitra2021}, which can be applied on top of existing recognition systems and do not require as much data as end-to-end solutions. 
Their VA-oriented approach was to optimize a small set of ASR decoder parameters on stuttered speech, such that the system is biased towards common VA phrases, and was effective in removing dysfluencies such as repetitions in speech.
One limitation of that work is that it was only applied to speech from 18 PWS, so in this paper we validate results on our much larger 91 person study. 

\subsection{Accessibility of Speech Recognition Systems} 
Voice interaction is used by a wide range of people with disabilities---such as the speech input that enables many people with motor disabilities to control computing devices (e.g.,~\cite{zhong2014justspeak, corbett2016can}), and as an inherently accessible input modality for many blind and low vision users (e.g., ~\cite{pradhan2018accessibility, abdolrahmani2018siri}). However, this work often assumes that speech input itself is accessible.

Researchers have captured speech input challenges encountered by people with dysarthria~\cite{ballati2018assessing,projecteuphonia}, deaf or hard of hearing people~\cite{fok2018towards}, and people with aphasia~\cite{roper2019speech}. 
For stuttering specifically, there has been relatively little human-centered work~\cite{clark2020speech}. A recent exception comes from Bleakley et al.~\cite{bleakley2022exploring}, who report on interviews and a three-week diary study of smart speaker use at home with 11 PWS. 
While participants had ``some success’’ in using the device, challenges included the device timing out, social pressure when using the device around other people, and difficulty with specific sounds (especially those that could interfere with the wake word, i.e., ``OK Google’’). Participants reported repeating and reformulating utterances to overcome these issues. While this study provides an initial understanding of how PWS use VAs, stuttering severity of the participants is not reported, and an objective assessment indicating how well the smart speakers recognized their speech is not provided. In contrast, our study examines speech input technology more broadly (e.g., including dictation), and includes both quantitative subjective data from a larger sample of n=61 PWS and system performance on stuttered speech samples collected from n=91 PWS. 

A few other human-centered projects have looked at supporting speech therapy for PWS and at managing stuttering in everyday life. Demarin et al.~\cite{demarin2015impact} and Madeira et al.~\cite{madeira2013building} both present apps to record stuttering experiences with the goal of increasing awareness and informing therapy decisions; small user evaluations included 2 and 5 PWS, respectively.
Focused on managing speaking situations, Fluent~\cite{ghai2021fluent} supports people in writing scripts that they will be more likely to read aloud fluently. The tool uses active learning to identify words that the user is likely to encounter issues with, and suggests alternative words that may be easier to speak; no user study has been reported. Finally, McNaney et al.~\cite{mcnaney2018stammerapp} employed a multi-stakeholder co-design process to design StammerApp, a mobile app that allows users to set goals, practice, and reflect on their experiences in especially challenging situations. While the project did not focus on speech input, it provides a qualitative understanding of a total of 39 participants’ experiences with stammering, including specific challenges (e.g., talking on the phone), and a general desire to be able to ``stammer openly and feel more confident and relaxed with their own voice.'' Ultimately, our goal is for PWS to have this freedom and confidence when interacting not only with other people, but with speech input technology as well.



\section{Speech technology survey for PWS}\label{sec:survey_results}

Our first goal is to understand the experience of speech technology for PWS, including how well existing systems are perceived to work, what challenges arise, and areas for improvements. 
We accomplished this by deploying a comprehensive online survey to 61 PWS in the United States during a three-month period in mid-2022.





\subsection{Survey Methods}

\subsubsection{Survey Design and Protocol}\label{survey_design}

Screened participants filled out the survey online using a commercial survey website. The survey took 30 to 60 minutes to complete and included the following four primary sections with up to 77 questions depending on conditional logic. 
\begin{description}
    \item[Background:] Demographics (age group, gender, race/ethnicity); self-description of speech characteristics; intelligibility of speech to family/acquaintances/strangers; ownership and usage of smart devices.

    \item[Voice Assistants (VAs):] Usage, usefulness, and accuracy of VAs; why and when do PWS use VAs; when do VAs break and areas for improvements; awareness and effectiveness of features including endpointing, hold-to-speak, invocation models, and other common features. 

    \item[Dictation Systems (DSs):] Usage, usefulness, and accuracy of DSs; why and when do PWS use DSs; when do DSs break and areas for improvements; awareness and effectiveness of relevant features, whether people revise or restate transcriptions upon error. 

    \item[Future Technology:] Interest in future forms of VAs/DSs, especially if speech were perfectly transcribed.
\end{description}

Many of these questions had multiple choice responses and open-ended text input to further explain responses. For questions related to the endpointer and other technical material, layperson descriptions like ``How often does a VA cut off your speech'' were used. See the Appendix for the list of survey questions discussed here. 

\subsubsection{Participants}\label{participants}
Participants were recruited through social media, referrals from speech-language pathology clinics, and word of mouth. 
To participate, individuals had to self-identify as having a stutter, after which they signed an Informed Consent Form and were screened by a speech-language pathologist (SLP) who listened to their connected speech patterns. 
Stuttering characteristics such as frequency, duration, and distribution of dysfluency types 
can vary greatly across individuals \cite{tichenor2021variability}, thus we aimed to recruit a broad range of participants with different stuttering severities. 
While there are many clinical severity assessments (e.g., \cite{SSI,yaruss2006overall}), 
we chose the Andrew \& Harris (A\&H) Scale~\cite{AndrewsHarris1964} which can be performed entirely using audio recordings. 
The SLP performed this assessment by examining dysfluency rates in audio recordings of a participant reading a passage, talking impromptu about a topic of their interest, and speaking with other individual(s).
The result was a grading of mild (0-5\% of words have dysfluencies), moderate (6-20\% of words), or severe ($>$20\% of words) stuttering.  

With the A\&H scale, 12 participants were rated as having a mild stutter, 31 as moderate, and 18 as severe. 
When asked about whether people understood their speech, almost all participants (93.4\%, n=57/61) reported that friends and close contacts `always' or 'usually' understand their speech, however this was lower for acquaintances (73.8\%, n=45) and strangers (67.2\%, n=41). 
Many individuals noted that their speech qualities change day-to-day (65.5\%, n=40) or throughout the day (55.7\%, n=34) and on top of typical dysfluencies such as blocks, prolongations, and repetitions, their voice may be strained (42.6\%, n=26) or breathy (16.4\%, n=10). 
Most individuals noted that speaking requires `some' physical effort (67.2\%, n=41) with fewer responses of `very high` (11.4\%, n=7) or `very low` (9.8\%, n=6).

In terms of demographics, there was an even gender split between men (50.8\%, n=31) and women (49.2\%, n=30), with no participants reporting another gender. Most participants were under 40 years old: aged 18-19 (n=2), 20-29 (n=28), 30-39 (n=21), 40-49 (n=3), 50-59 (n=4), 60-69 (n=2), and 70-79 (n=1). Individuals identified as white (n=38), black or African American (n=12), Hispanic (n=5), Asian (n=2), American Indian or Alaska Native (n=1), or preferred not to answer (n=3). In terms of technology use, all participants owned smartphones, with most having an iPhone (93.4\%, n=57) and four having an Android phone. 
Over half (59.0\%, n=36) had smart speakers in the home: 27 had an Amazon device, 5 had a Google device, and 4 had an Apple device. For other technology, 58 participants reported having a computer, 35 had a tablet, and 27 had a smartwatch or fitness tracker.

\subsubsection{Analysis}
For closed-ended questions, we report primarily on descriptive statistics including counts and proportions. We have included a few comparative statistical tests for high-level questions of how often participants use VAs vs. dictation, and whether there is an impact of stutter severity ratings on usage. In doing so, we selected tests that are appropriate for ordinal data: Mann Whitney U and Kruskal Wallis tests. For open-ended questions, we conducted two open coding passes on each question and report on the prominent themes identified in this analysis to help contextualize the quantitative results.

\subsection{Survey Findings}
We present survey results including an overview of speech interaction and usage, challenges related to both VAs and dictation, and factors beyond technical considerations that could impact future adoption.

\subsubsection{Speech Interaction Usage and Utility} 
\begin{figure}[h!]
    \centering
    \subfloat[\centering ]{{\includegraphics[width=0.35 \paperwidth]{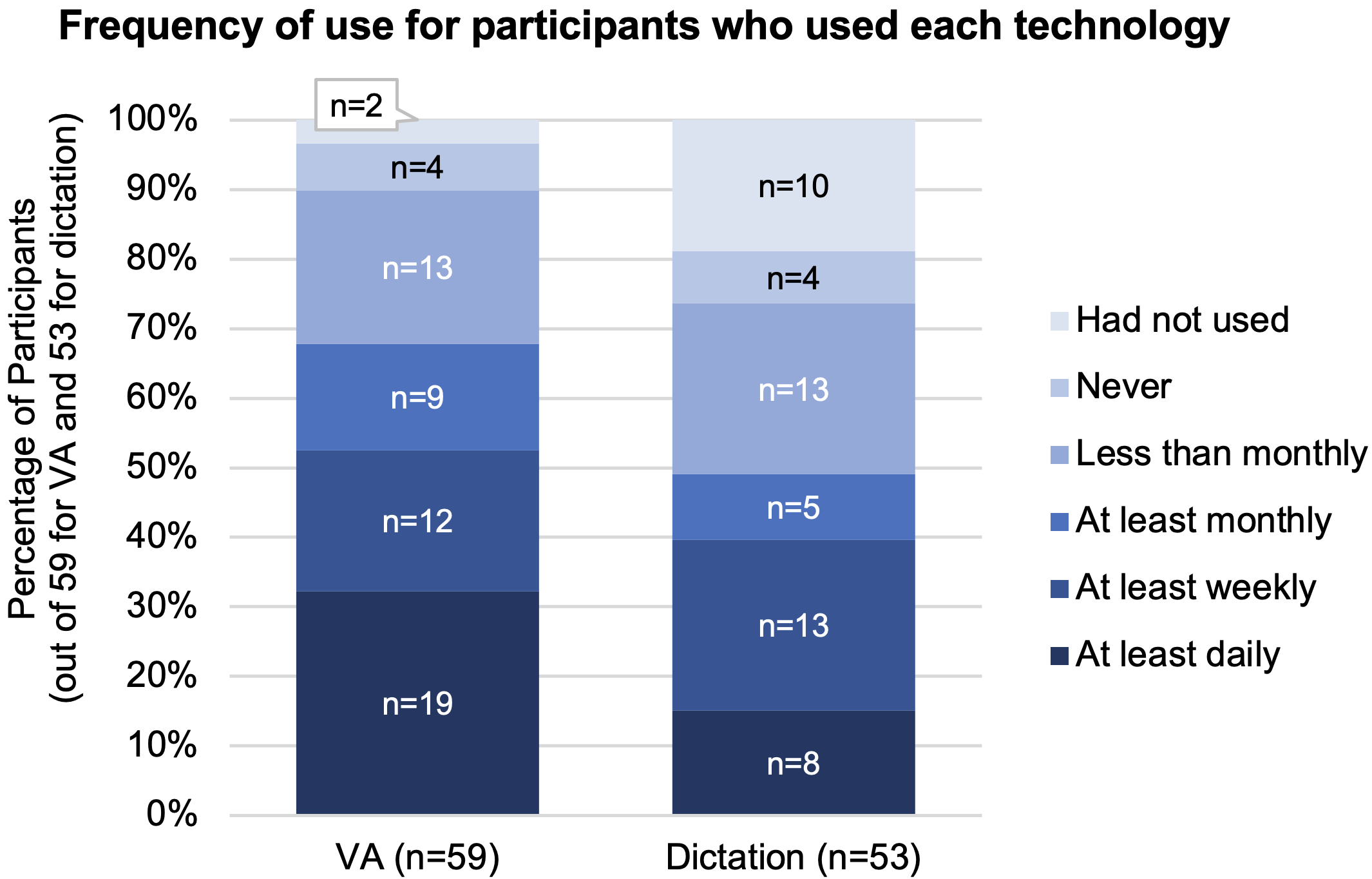} }}%
    \qquad
    \subfloat[\centering ]{{\includegraphics[width=0.35 \paperwidth]{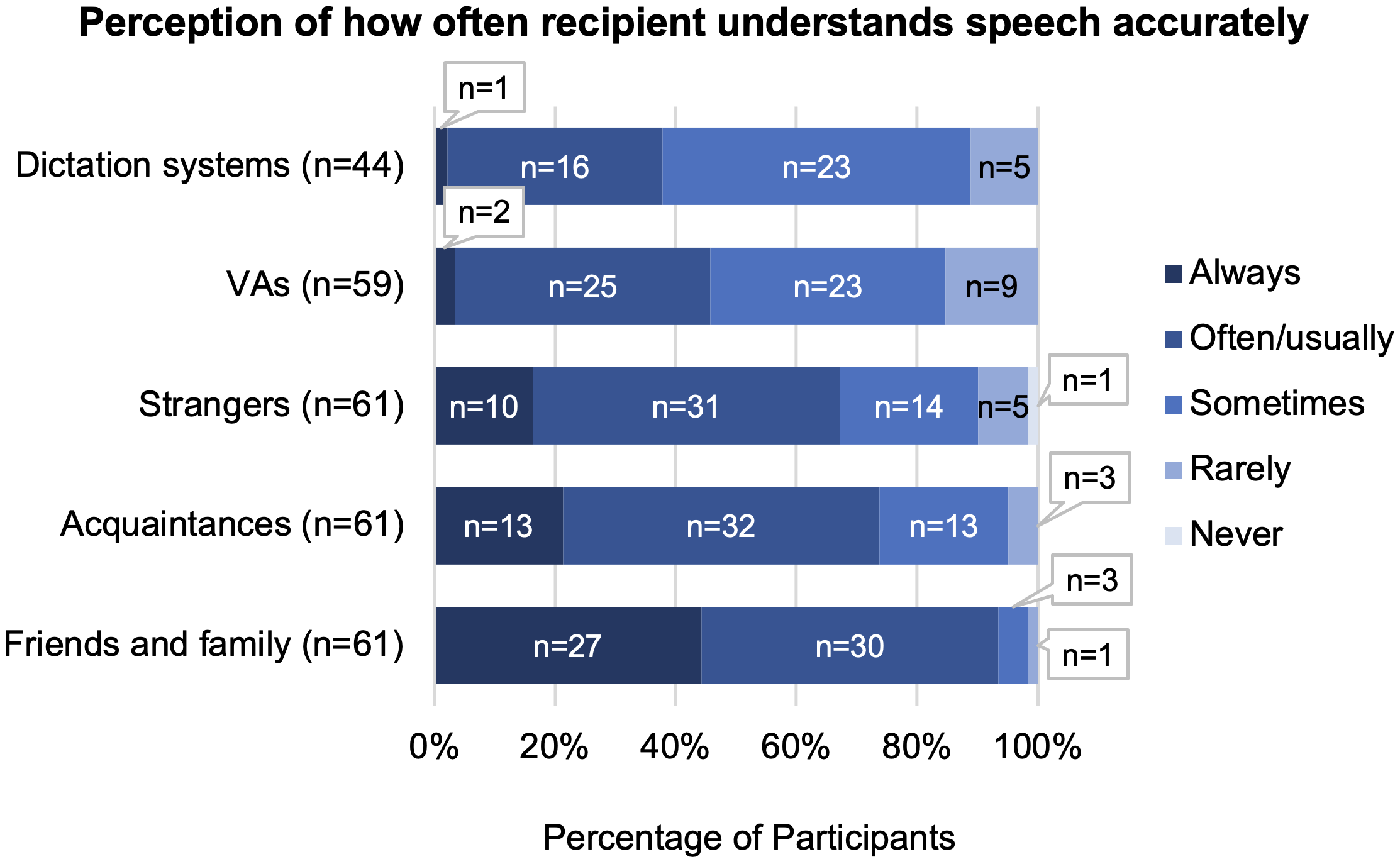} }}%
    \caption{(a) Frequency with which participants \textit{who were familiar with} VAs and/or dictation use each technology, showing significantly greater use of VAs; two participants who responded ``I don't know'' are not included. (b) Frequency with which participants feel different recipients, including VAs, dictation, and other people, understand their speech accurately. For dictation and VAs, this analysis includes participants who had reported experience using each technology (n=44 for dictation because one participant did not answer this question). Overall, for both VAs and dictation, the majority of participants felt these systems understood their speech at best `sometimes,' in contrast to much higher comprehension rates when speaking to other people.}%
    \label{fig:freqUseAndPerceivedAccuracyVAandDictation}
    \Description{
    Figure (left): Bar charts for frequency of usage for VA and Dictation, ranging from had not used to daily use. See text in the paper for percentage breakdowns. 
    Figure (right): Bar charts for perception of how often recipient’s think systems (VA or Dictation) or people (strangers/acquaintances/friend & family) understand their speech. These rang from Never to Always. People tend to think that VAs understand their speech a little more often than Dictation (see paper for details). For around 93\% of PWS, friends and family always or often/usually understand them, which decreases to 74\% for acquaintances and around 67\% for strangers. In contrast, responses for sometimes, and rarely have similar patterns increase, meaning that PWS do not think they are understand as well by people they are less familiar with. The percentage of participants who are always or often/usually understood by VAs is around 46\% and this number is 39\% for
    Dictation.
    }    
\end{figure}

Most participants were familiar with speech recognition systems (VA: 100\%, n=61; DS: 90.2\%, n=55), although more had used VAs (96.7\%, n=59) than dictation (73.8\%, n=45).
Overall, among participants who were familiar with the given technology, VAs were used significantly more often than dictation (Mann-Whitney U test\footnote{We used a Mann Whitney U test because this is ordinal data and the data was not fully paired---not all participants who had used VAs had also used dictation. This analysis excludes the two participants who answered ``I don't know'' for their usage of VAs and dictation. Participants who were familiar with but had not used a given technology were combined with those who had used it but reported their current usage frequency as ``never,'' as shown in Figure~\ref{fig:freqUseAndPerceivedAccuracyVAandDictation}.}: $U=1937$, $n_1=59$, $n_2=53$, $p=.019$).
As shown in Figure~\ref{fig:freqUseAndPerceivedAccuracyVAandDictation}, about half of the participants familiar with VAs used one on a daily (32.2\%, n=19/59) or weekly basis (20.3\%, n=12/59),
whereas fewer used dictation on a daily (15.1\%, n=8/53) or weekly basis (24.5\%, n=13/53). 
This VA usage rate is somewhat lower than recent estimates for the general population in the US that suggest 57\% use VAs daily and 23\% do so weekly \cite{npr_smart_audio}, perhaps reflecting the accessibility barriers faced by PWS. 

\begin{figure}[h!]
    \centering
    \subfloat[\centering ]{{\includegraphics[width=0.4
    \paperwidth]{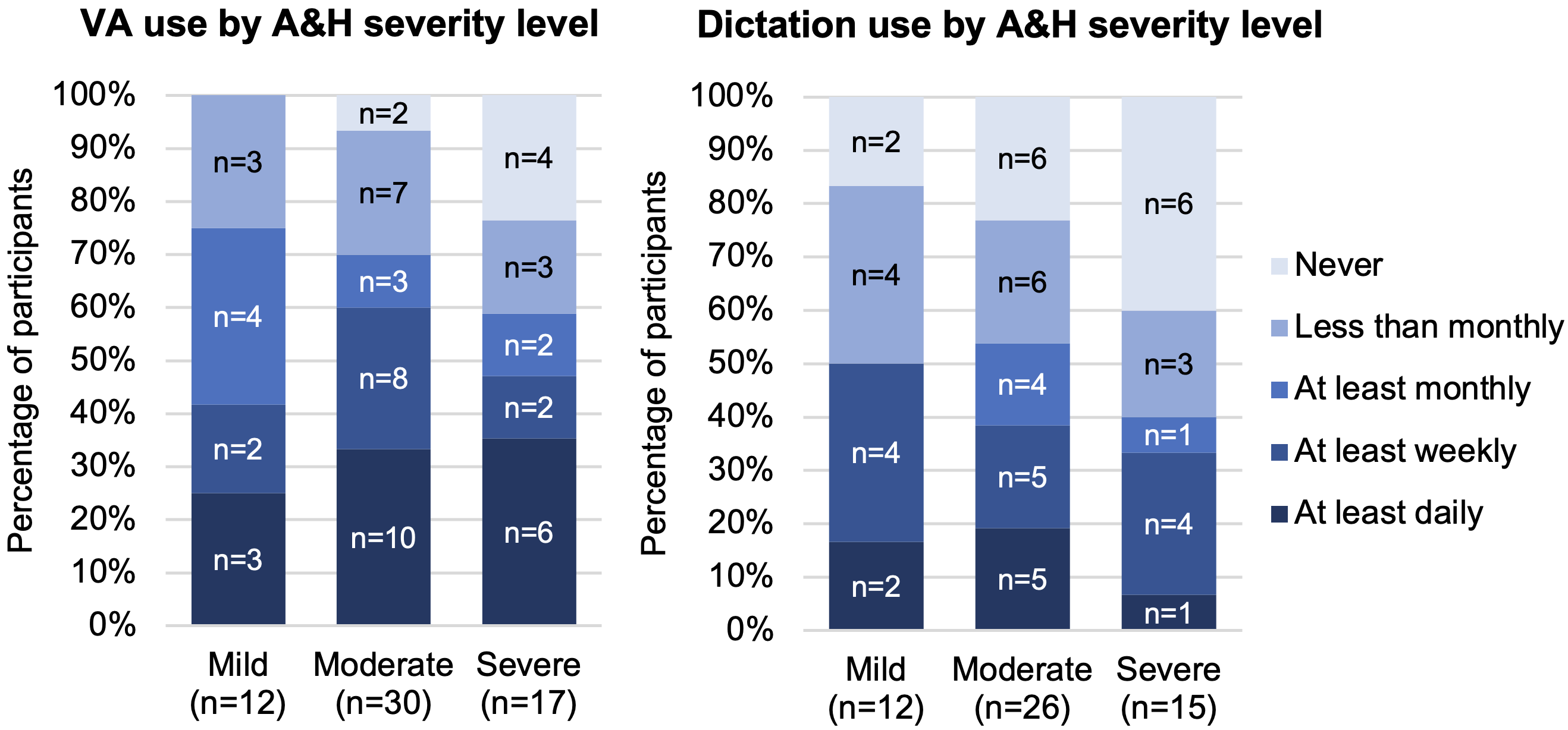} }}
    \qquad
    \subfloat[\centering ]{{\includegraphics[width=0.30 \paperwidth]{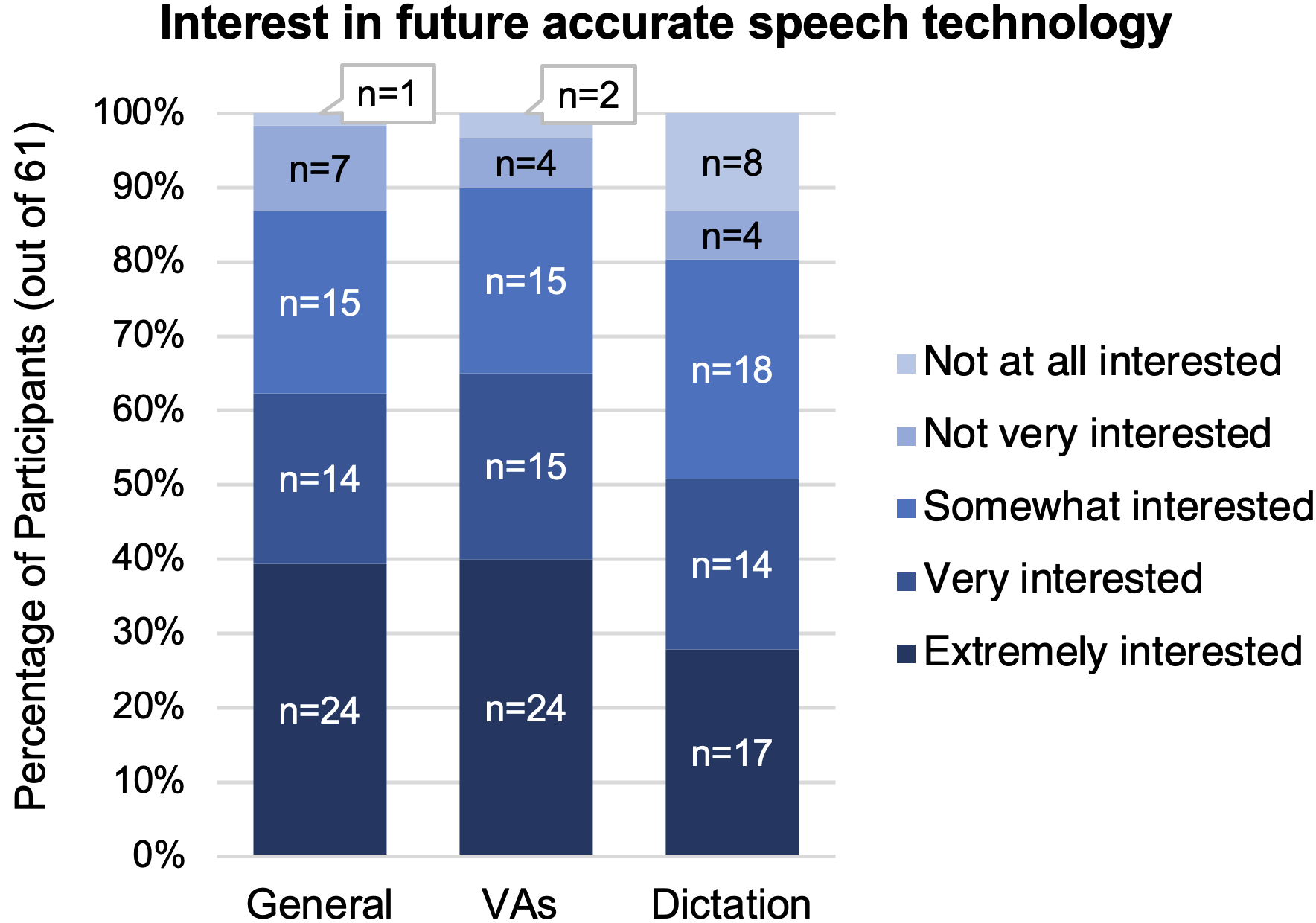} }}
    \caption{(a) Dictation and VA usage frequency based on A\&H severity ratings for participants \textit{who were familiar with} VAs (left) and those \textit{who were familiar with} dictation (right), showing varied usage frequency regardless of graded severity. Most respondents have reported use of VAs (n=59) and dictation  (n=53). Two respondents who said answered ``I don't know'' for VA and dictation frequency were not included in this analysis. 
    (b) Interest levels in future hypothetical speech technologies that can understand speech with dysfluencies accurately (n=61), showing high interest in general speech technology, as well as VAs and dictation.
    }
    \label{fig:usage_frequency_by_severity_future_tech}
    \Description{
Figure (left): Bar charts for “VA use by A&H severity” and “Dictation use by A&H severity” with values of Never/less than monthly/more than monthly/at least weekly/at least daily, broken down by mild/moderate/severe participants. There are a higher percentage of people with severe A&H severity ratings that chose ‘never’ than moderate, and more moderate that chose ‘never’ than mild. 
Figure (right): Bar charts for “Interest in Future Accurate Speech Technology” ranging from Not at all/not very/somewhat/very/extremely. See the paper for breakdowns, but generally people at least 50\% are extremely or very interested.
    }
\end{figure}

Because past work has shown that stuttering severity affects speech recognition accuracy~\cite{mitra2021}, we further examined the relationship between A\&H severity ratings and reported dictation and VA usage. These results, shown in Figures~\ref{fig:usage_frequency_by_severity_future_tech} indicate no clear patterns based on graded severity. Kruskal Wallis tests did not reveal significant impacts of A\&H severity level (i.e., mild, moderate, and severe) on reported frequency of use for either VA usage or dictation usage.\footnote{A Kruskal Wallis test is a non-parametric equivalent to a one-way ANOVA, and is appropriate for ordinal data such as our frequency of use ratings (i.e., 1-At least daily, 2-At least weekly, 3-At least monthly, 4-Less than monthly, 5-Never). We conducted two separate Kruskal Wallis tests to examine the effect of the independent variable of A\&H severity level (mild, moderate, or severe) on the dependent variable of frequency of use for each of VA and dictation use. The tests were not significant at $p < .05$ for both VAs ($H(2)=0.57$, $p=.752$) and dictation ($H(2)=1.53$, $p=.465$). These analyses were conducted on $n=57$ participants (VA) and $n=43$ (dictation) who had reported experience with the given technology and frequency of use.} This lack of significance may be due to low statistical power related to our sample size and variability in how people rate their frequency of use, but we return to additional considerations in Section~\ref{sec:discussion}.

Turning to \textit{how} and \textit{why} participants adopted VAs and dictation, we first examined what platforms they used.
Among people with experience using VAs or dictation, phones were the most common platform (VA: 84.7\%, n=50/59; DS: 88.9\%, n=40/45) followed by smartwatches (VA: 25.4\%, n=15; DS: 24.4\%, n=11) and car infotainment systems (VA: 25.4\%, n=15; DS: 20.0\%, n=9). Smart speakers were also relatively common for VA use (47.5\%, n=28) but not used as much for dictation (9.0\%, n=4)---which is expected given that smart speakers are often communal devices and have fewer dictation use cases. Ten or fewer participants had used speech on tablets or traditional computers for both VAs and dictation.


For why they use speech technology, a majority of participants find speech recognition systems to be at least ``somewhat'' useful; see Table~\ref{tab:perceived_utility}a for detailed utility ratings.
For those surveyed that use VAs, common use cases generally mirror those from the general population~\cite{AmmariCHI21}. 
For VA users (Table~\ref{tab:perceived_utility}b), participants most commonly reported listening to music, making phone calls, asking general questions, checking the weather, setting alarms and timers, and sending messages.
For dictation users (Table~\ref{tab:perceived_utility}c), the common use cases were writing text messages, taking notes, writing email, and dictation within a web browser.
However, about a quarter of participants familiar with each of the two speech technologies felt that they were `not very' 
or `not at all' 
useful.
Overall, reasons were split between participants who saw no inherent value in VAs (e.g., \textit{``I haven't felt a need for them in my life''} [P2-51]) and those who encountered accessibility challenges due to stuttering (e.g., \textit{``[...] I often have to repeat myself so many times for it to understand me that it’s quicker and easier for me to just use my hands [...]''} [P2-64]). 
We examine these and other challenges in more detail in the following sections.

\begin{table*}
\begin{subtable}[]{0.33\textwidth}
    \begin{tabular}{|l|cc|cc|}
    \hline
    & \multicolumn{2}{c}{\textbf{VAs}} & \multicolumn{2}{|c|}{\textbf{Dictation}} \\
    \textbf{Utility } & \textit{n} & \textit{\% of 59} & \textit{n} & \textit{\% of 45} \\
    \hline
    Extremely & 6 & 10.2\% & 4 & 8.9\% \\
    Very & 18 & 30.5\% & 14 & 31.1\% \\
    Somewhat & 19 & 32.2\% & 17 & 37.8\% \\ 
    Not very & 9 & 15.3\% & 9 & 20.0\% \\
    Not at all & 7 & 11.9\% & 1 & 2.2\% \\
    \hline
    \end{tabular}
    \caption{} \label{capA}
\end{subtable}
\hfill
\begin{subtable}[]{0.32\textwidth}
    \begin{tabular}{|l|cc|} \hline
    \textbf{VA Uses}        & \textit{n} & \textit{\% of 59}  \\ \hline
    Music                        & 41         & 69.5\%              \\
    Phone calls                   & 41         & 69.5\%              \\
    General Q\&A            & 39         & 66.1\%              \\
    Weather                      & 36         & 61.0\%                \\
    Alarms                       & 31         & 52.5\%              \\
    Sending messages & 31 & 52.5\% \\
    Timers                       & 30         & 50.8\%              \\ \hline 
    \end{tabular} 
    \caption{} \label{capB}
\end{subtable}
\hfill
\begin{subtable}[]{0.25\textwidth}
    \begin{tabular}{|l|cc|} \hline                             
    \textbf{Dictation Uses} & \textit{n} & \textit{\% of 45} \\ \hline 
    Texting                      & 42         & 93.3\%              \\
    Notes                        & 17         & 37.8\%              \\
    Email                        & 11         & 24.4\%              \\
    Web browser                    & 10         & 22.2\%             \\ \hline 
    \end{tabular}
    \caption{} \label{capC}
\end{subtable}
\caption{(a) Perceived utility ratings from survey participants who had experience with VAs and dictation. Most found VAs and dictation to be at least ``somewhat'' useful, although about a quarter found them to be ``not very'' or ``not at all'' useful. (b and c) The most common use cases reported by at least 50\% of participants who had experience with VAs and by at least 20\% of participants who had experience with Dictation.  
}
\label{tab:perceived_utility}
\end{table*}

\subsubsection{Voice Assistant Challenges}
To understand the range of challenges PWS face when using VAs, we asked about what factors prevented participants who were familiar with VAs from using them more often. The most common were `voice assistants cut off my speech before I finish talking' (61.0\%, n=36/59) and `voice assistants don't understand my speech well enough' (57.6\%, n=34). Indeed, as shown in Figure~\ref{fig:freqUseAndPerceivedAccuracyVAandDictation} (b), a majority of participants (54.2\%, n=32) felt VAs only recognized their speech `sometimes' or worse. As a result, many participants (37.3\%, n=22) reported `often' or `always' attempting to revise what they say due to inaccuracies in speech recognition. 
Other concerns, however, were not purely technical. 
Twenty-two participants (37.3\%) reported not wanting other people to hear them talking to the VA, reflecting a social consideration that past work has also identified~\cite{bleakley2022exploring}. 
As well, many participants (22.0\%, n=13) cited that it takes `too much physical effort for me to speak.' 
Only six participants (10.2\%) responded that perceived utility of voice assistants prevented them from using the technology more often. 

Concern with endpointing---being cut off too soon---is typically more relevant to VAs than dictation, and was identified as a challenge in Bleakley et al.'s~\cite{bleakley2022exploring} interviews on VA use with 11 PWS. Our data demonstrates the pervasiveness of this problem: almost half of participants stated that VAs cut them off `always' (3.3\%, n=2) or `often' (37.7\%, n=23), while an additional 31 participants (50.1\%) reported being cut off `sometimes' or `rarely.' For the 57 participants who had at least some experience being cut off, when asked if they knew why it happened, by far the most common explanation was that there was a pause or block in their speech and the system thought they had finished speaking (77.2\%, n=44 of 57). 
Being cut off could also have consequential impacts, such as a poor user experience described by [P2-42]: \textit{``The [smart speaker] which I use most often does not like any hesitations in speech. It assumes the slightest pause means the person is done speaking. So I get cut off often. It creates a sense of having to rush which makes my speech worse.''} And, as [P2-77] described, impacts can also be emotional: \textit{``I block a lot on my name and vowel words so when I take time to talk it is frustrating when they cut me off [...] I have to deal with people often trying to speak for me and having a voice assistant doing that because I can't have one who understands my speech rhythm makes me not want to use it even more.''}
We address endpointing errors in Section~\ref{sec:endpointer} and significantly reduce the problem on our dataset of speech samples.

Some smartphones (e.g., Apple iPhone) offer an alternative to automatic endpointing, where the user can push down a button (physical or virtual) while speaking and only release it when done.
Twenty-seven participants (44.3\%) reported using a device that supports this functionality. Of those 27, almost all found being able to press a button and speak to be useful: 51.9\% (n=14 of 27) `extremely useful,' 18.5\% (n=5) `very useful,' and 29.6\% (n=8) `useful.' Reasons provided for this utility were most commonly that it eliminates being cut off, but some participants also explicitly referred to a psychological impact of feeling less rushed or worried, such as: \textit{``I can fully get my phrase in without worrying about blocking''} [P1-62] and \textit{``I don't feel rushed to finish what I need to say''} [P1-45]. 

Another potential issue unique to VAs is invocation.
Most VAs use a wake word like ``Alexa,'' ``OK Google,'' or ``Hey Siri'' for invocation, which can be difficult for some PWS to use. Almost half of participants reported that these wake words `always' (11.5\%, n=7) or `often' (36.1\%, n=22) worked for them. 
But, problematically, 17 participants (27.9\%) reported wake words only working `sometimes' and 8 participants reported `rarely' or `never' (13.1\%); the remaining participants had not used a wake word. 
While we did not ask participants to elaborate on their use of wake words, a handful of responses to other open-ended questions referred to challenges with wake words, including [P2-1] who stated \textit{``Problem with wake words is my biggest frustration. I would want to change it to something I can say easily''}. 
While popular consumer systems do not allow for arbitrary wake words, some offer multiple options (e.g.,
Alexa has at least five options and Google Assistant has two). 
Additionally, a few participants mentioned that pressing a button to talk (as described above) eliminated issues with wake words by allowing an alternative explicit signal that the user is ready to talk. [P1-28] noted \textit{``It's [Hold/Squeeze to Speak is] useful on my phone to bypass the lag that occurs when [the VA] does not recognize me using the wake phrase''}, and [P1-73] uses it \textit{``Because it knows when to start listening and when I'm done.''}


\subsubsection{Dictation System Challenges} 
As with VAs, we aimed to broadly understand what factors impact dictation adoption for PWS. When asked what reasons prevent them from using dictation more often, 56.4\% (n=31) of the 55 participants who were familiar with dictation noted that `dictation tools don't understand my speech well enough.' Of those who had used dictation, a majority (59.1\%, n=26 of 44 who answered this question) felt the ASR only recognized their speech accurately ``sometimes'' or less (Figure~\ref{fig:freqUseAndPerceivedAccuracyVAandDictation}, right). As a result of these errors, participants reported attempting to revise or restate what they say, with about half of participants doing so `always' (6.7\%, n=3/45) or `often' (42.2\%, n=19), and an additional 35.6\% (n=16) doing so `sometimes.'
Finally, and similar to VAs, participants also commonly cited non-technical reasons for not using dictation more, most notably the physical effort it takes to speak (32.7\%, n=18 of the 55 familiar with dictation) and not wanting other people to hear them dictate to the device (30.1\%, n=17). Additionally, 11 participants (20.0\%) reported that low utility prevented them from using dictation more often, and 12 (26.7\%) felt that they use more refined language when they type than when they speak. 

In contrast to VA systems, which predict the most likely command based on a spoken phrase, with dictation the user's speech is transcribed verbatim. This difference is reflected in participants' experiences with dictation errors. 
When asked an open-ended question about what types of errors tend to occur with dictation, by far the most common response, mentioned by a majority of participants, was errors in the ASR transcription (e.g., \textit{``Message contains different words than I intended''} [P2-5]). 
Some participants (13.3\%, n=6) also mentioned being cut off, although this was a much less common error than with VAs because many dictation systems do not use an automatic endpointer. 
A few participants mentioned poor performance with punctuation, proper nouns, and slang. 
Finally, some participants mentioned specific stuttering characteristics and challenges, including their repetitions being displayed in the transcribed text
(\textit{``Misunderstood due to repetition or actually repeats word''} [P2-46]), the impact of blocks 
(\textit{``words that I have a block in the middle of will turn into a strange phrase''} [P2-51]), and interjections being shown, as described by [P2-77]: \textit{``because I have blocks and use filler words, the fillers are listed on the text message all my disfluencies are written down and when I read over the text it makes no sense and I have to type it myself instead because I will never be fluent all the time.''}

\subsubsection{Beyond Technical Considerations} 
The findings above suggest that technical improvements in invocation, endpointing, and ASR accuracy should be welcomed by many PWS and ultimately increase adoption of speech technologies by this population. However, future adoption also depends on factors beyond technical improvements, 
such as social concerns and the physical difficulty of speaking, emphasizing that speech input may not be relevant to all users even with substantial accuracy improvements. 


To begin to understand what future adoption might look like,
we asked participants to imagine speech technologies that can accurately understand their speech. 
Within this context, most participants were at least `somewhat` interested in VAs (86.9\%, n=53/61): `extremely’ (n=25), `very' (n=13), `somewhat' (n=15), `not very' (n=7), `not at all' (n=1) and similar for dictation (Figure~\ref{fig:usage_frequency_by_severity_future_tech}, right). 
Positive responses reflected earlier feelings about VAs and dictation, notably increasing efficiency, ease and convenience, being able to do actions hands-free (including while driving), and generally having confidence in the system. 
For example, [P1-66], who currently uses VAs at least monthly and has not tried dictation wrote, \textit{``I would be extremely interested in this, especially knowing it's something I don't have to worry about the speech input being able to understand me. It would remove my hesitancy and frustration with my stuttering from the equation.''}


However, the negative and neutral reactions to this future technology scenario indicated a preference for typing, seeing no need for speech input, and value depending on context (e.g., \textit{``It would depend on the device''} [P1-64], who uses VAs weekly and dictation less than monthly). 
This preference for typing by some participants also arose in other responses throughout the survey, reflecting in some cases the physical effort required to speak. 
For example, \textit{[P2-73] said, ``It's a nice thought but it's still easier for a person who stutters to type most of the time. Unless our hands are filthy or tied up''} and \textit{[P2-87] ``It still requires effort to speak even if it understand me perfectly, and I am good at controlling them manually.''} 
In addition, 9.8\% (n=6) stated  `voice assistants don't seem useful.' 
while [P2-51] said, \textit{``I'm not sure. As someone who stutters, and just situationally in my life, I can't imagine dictation regularly being a more convenient option than typing.''} 

Additionally, social factors may influence future adoption, a consideration touched on in past work with PWS~\cite{bleakley2022exploring}. 
While the survey did not explicitly ask participants about the social context of speech technologies, as already noted, many participants felt that having to speak in front of other people limited their willingness to use speech technology. For example, [P2-1] wrote \textit{``[...] I would be reluctant to use voice commands in public because of my speech impediment''}, and [P1-58] wrote, \textit{``I still would not use it at work. I would use it in spaces where I am alone, with my spouse or family.''} 
Some participants also commented on how anxiety affects their stuttering, which translated to speech technology use in different ways. [P2-88], for example, expressed that \textit{``...before I am about to use voice recognition I get a bit nervous because I think that I have to speak without stuttering for it to work''}---a concern that could be alleviated with improved ASR. 
At the same time, some participants expressed that speaking to devices  reduces anxiety compared to speaking to other people, including: \textit{``I don't stutter much when talking to myself so I usually don't experience this problem [dictation errors]''} [P2-15] and \textit{``I usually stutter in high anxiety situations not when I'm by myself''} [P2-70]. Depending on the individual person and the context of use, these social factors could prove to be barriers to adoption.


\subsection{Summary}
This survey shows that many PWS are using speech technology on a regular basis, albeit at lower rates than the general population (at least for VAs). However, accessibility challenges include being cut off or misunderstood by VAs, and having stuttering events such as repetitions and prolongations result in transcription errors with dictation. Many participants were interested in using future technologies that could recognize stuttered speech accurately, but other responses suggest that a subset of PWS would still choose not to use such technology due to the physical effort of speaking and social considerations.

\section{Speech Recognition Technology: Data \& Experiments}\label{sec:asr_results}


Building on findings from the survey, we turn to recorded speech data from 91 PWS (53 of whom also completed the survey). We quantify how dysfluencies in speech from PWS manifest in the types of recognition errors encountered by survey participants. We also investigate changes to three stages of the speech pipeline introduced in Section~\ref{sec:rw-speechrecsystems} that address issues identified in the survey. 
We focus on interventions that build on pre-trained production-grade models, as oppose to training from scratch, so we can optimize on relatively small amounts of data from PWS: 
\begin{enumerate}
    \item \textit{Tuning the endpointer model} using speech from PWS, to reduce speech truncations (cut-off speech).
    \item \textit{Tuning the decoder in the ASR model} using speech from PWS, to increase ASR accuracy. 
    \item Applying posthoc \textit{dysfluency refinements} to ASR transcriptions to remove dysfluencies such as repeated words.
\end{enumerate}

\subsection{Speech Data Collection}\label{sec:speech_data}

Participants were asked to record themselves speaking 121 VA commands and up to 10 Dictation phrases that they could imagine saying to a smart speaker or device.
An initial cohort of 50 participants (Phase 1: 25 mild; 18 moderate; 7 severe)\footnote{A\&H severities were computed using the same process as with the speech survey.} was collected starting in Summer 2020 and 41 additional (Phase 2: 0 mild; 26 moderate; 15 severe) in 2021.\footnote{65 participants successfully recorded 131 utterances, 2 recorded 130 utterances, and 24 only recorded 121 VA commands.} 
VA phrases were based on prompts where participants came up with their own utterance to accomplish a given task. 
This was designed to prevent participants from simply reading pre-defined phrases in a monotone manner, 
and some examples include ``Find out when a place you might want to visit will open,'' ``Find the lyrics to a song that you like,'' and ``Ask a question about something that is not well known.''
Dictation phrases consisted of fake text messages that they might send to a friend. 
Audio clips were captured using a smartphone on their own in a quiet environment where participants were asked to speak naturally. 
The median utterance length was 6 words. 


All audio clips were manually transcribed with both the articulated phrase with dysfluencies (e.g., ``wh-(at) wh-(at) what is [...] the weather'' and the phrase that annotators believed the participant intended to say (e.g., ``what is the weather''). For each phrase, one human transcriber created an initial transcript and a second person reviewed the transcript; additionally, a random ~15\% of transcripts were spot checked by a third independent reviewer for quality control and were redone if needed.
For the 41 participants in Phase 2, which has a larger percentage of those with moderate or severe dysfluency rates, the transcripts were annotated with detailed dysfluency labels.
Specifically, for each phrase, an annotator segmented each dysfluency event by marking the start and end frames, and added one of the following five dysfluency labels: part-word repetition (e.g., ``ha- ha- ha- happy''), whole word repetition (e.g., ``I I I''), sound prolongation (``wasss''), blocks (audible and inaudible), and interjections (e.g., ``uh'', ``um'').
These annotations also went through a similar QA process where dysfluencies with incorrect timestamps or labels were marked as wrong. 

Across all clips with dysfluency annotations,  58.6\% of utterances contain one or more part-word repetitions, 5.5\% contain one or more whole word repetitions, 36.4\% prolongations, 38.7\% blocks and 2.8\% interjections.
Interjections were very common with one participant but less prevalent more broadly.
The rarity of interjections may be due to the short utterance lengths.
In general, the distribution of dysfluency types varies greatly on a per-participant basis. Some individuals frequently block but have few repetitions whereas others have frequent repetitions but few blocks. 
We look more at connections between dysfluencies and WER throughout our speech recognition experiments.

\subsection{Speech Recognition Results}

In this section we present baseline performance for both endpointer and ASR models on our dataset, as well as results from the three interventions that address issues that arose in the survey: (1) endpointer tuning to address the problem of being frequently cut off by VAs, (2) ASR decoder tuning to improve the ability to understand stuttered speech, and (3) refining dysfluencies in the transcribed speech that is output from the ASR model to improve dictation experiences for PWS. We also place these findings in historical context by investigating how ASR performance on speech from PWS has changed over the past five years, using archived consumer-grade ASR models that were publicly available from 2017-2022.

The baseline models are from the 
Apple Speech framework~\cite{SpeechFramework}, which uses a hybrid deep neural network architecture for the ASR system.
See \cite{huang2020sndcnn} for ASR model details, which at a glance is composed of an acoustic model, a language model, and a beam search decoder. 
The acoustic model maps audio to phone- or word-like intermediate representations, a language model encodes the probability of word sequences and acts as a prior over what words or phrases someone may have said, and a beam search decoder that 
efficiently computes candidate transcriptions.

\subsubsection{Endpointer Model Performance} \label{sec:endpointer}
An endpointer model identifies when the user stops speaking, and must balance the desire for a low truncation rate (i.e., what percent of utterances are cut off too early) with the desire for a minimal delay after speech (i.e., time from the end of the utterance to when the VA stops listening). 
Our base model is trained on completed utterances from the general population, and predicts the end of a query using both auditory cues like how long the user has been silent as well as ASR cues like the chance a given word is the final word of an utterance.
For each input frame (time window), the model outputs the likelihood of utterance completion. 
Once the output exceeds a defined threshold, the system stops listening and moves on to the next phase of processing.

\begin{figure}[t!]
    \centering
    \subfloat[\centering ]{{\includegraphics[width=0.33 \paperwidth]{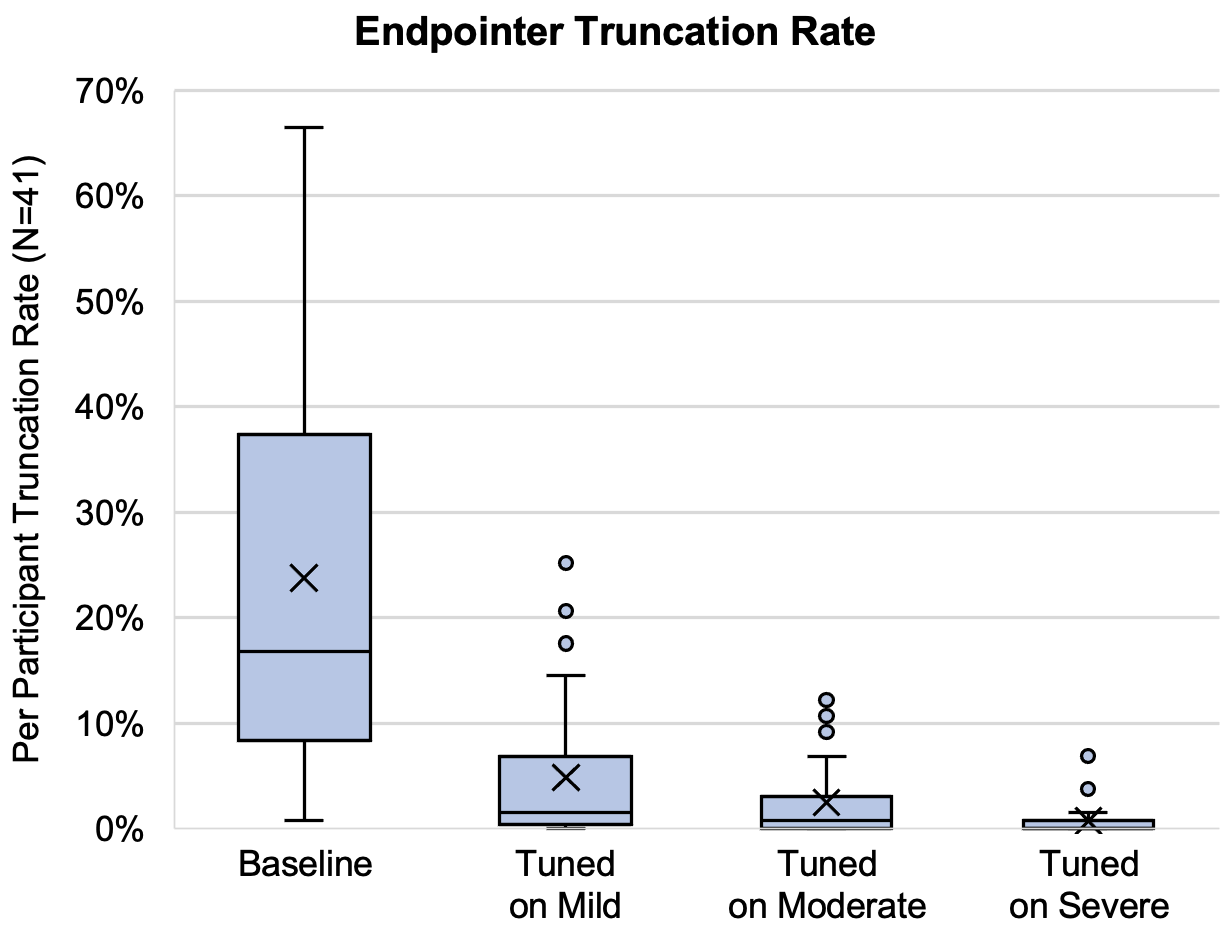} }}%
    \qquad
    \subfloat[\centering ]{{\includegraphics[width=0.33 \paperwidth]{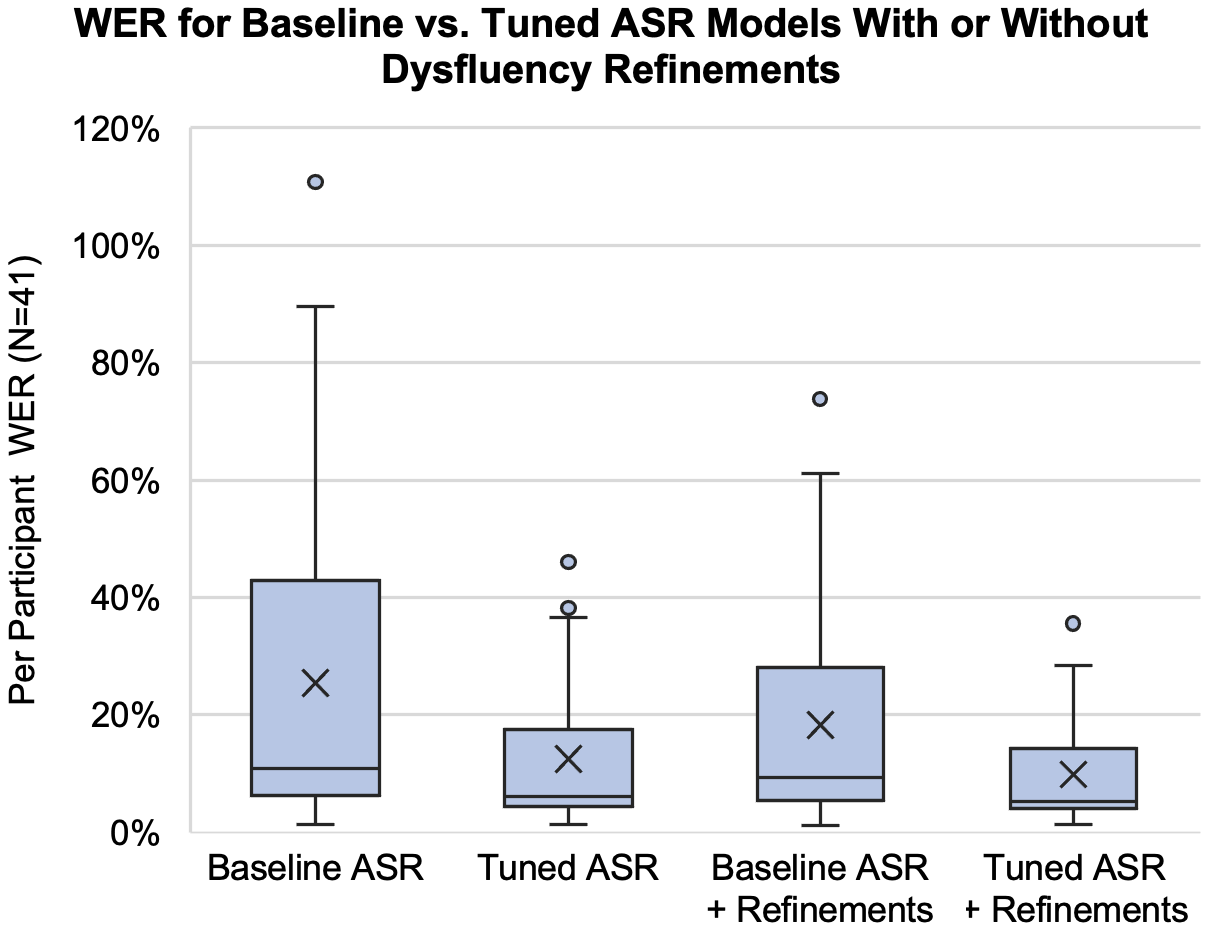} }}%
    \caption{(a) Endpointer and (b) ASR performance for Phase 2 participants (N=41). In both cases, the tuned models are trained on Phase 1 data. The $\times$ symbol marks the mean in each box, whereas the solid internal line is the median. (a) Truncation rates are substantially lower than the baseline for all tuned thresholds, and the moderate threshold results in a rate of less than the 3\% targeted for the general population.  
    (b) For WER, the ASR decoder tuning and the dysfluency refinements significantly improve performance compared to the baseline, and the combination of the two results in a further significant improvement.  
    }%
    \label{fig:boxplots_endpointer_and_wer}
     \Description{
    Figure (left): Box plot for endpointer truncation rate, with four boxes each representing baseline endpointer model and tuned models on speech data from participants with mild, moderate and severe A&H ratings. The tuned models reduce truncation rates from about 24\% to below 5\%.   
    Figure (right): Box plot for ASR performance (WER), with four boxes representing baseline ASR and tuned ASR models with or without dysfluency refinements. The ASR decoder tuning improves performance significantly,  for the third quartile. The dysfluency refinements further improve performance compared to the baseline for both the baseline and tuned ASR.
    }  
\end{figure}

\paragraph{Baseline endpointer.} The baseline likelihood threshold from the model we used was set such that 97\% of utterances from the general population data are endpointed correctly (i.e., truncation rate of 3\%). When evaluated on our data from PWS, which is likely poorly represented in general population data, a much higher portion of data is truncated early. 
Across the 41 Phase 2 participants,\footnote{For consistency across later experiments where we use Phase 1 data to tune models, we only report performance here for Phase 2 data.} the baseline endpointer model truncates on average 23.8\% of utterances ($SD$=19.7, $Median$=16.8, $IQR$=29.0); see Figure~\ref{fig:boxplots_endpointer_and_wer}(a). Truncation rates also vary substantially per person---for 7 of 41 participants over 50\% of utterances are cut off early. This high truncation rate reflects the survey findings, where a majority of participants reported that early truncation was a key issue.

Our hypothesis, based on the literature and understanding of stuttering, is that blocks, which are often expressed as inaudible gasps, cause most endpointing errors. 
We validated this using the dysfluency annotations by computed Spearman's rank correlation\footnote{\label{spearman_footnote}Spearman's rank correlation is used to correlate variables that are not distributed by a Gaussian distribution 
such as our data that is highly skewed.}
and found truncation rates are significantly positively correlated with rates of blocks ($r(39)=.64$, $p<.001$), part-word repetitions ($r(39)=.44$, $p=.004$), and interjections ($r(39)=.48$, $p=.002$). Correlations with prolongations ($r(39)=.17$, $p=.301$) and whole word repetitions ($r(39)=.18$, $p=.271$) are not statistically significant.

\paragraph{Tuned endpointer threshold (Intervention \#1)}
To improve the VA experience for PWS, we investigate performance of three new thresholds that could reduce truncations for people with different rates of dysfluent speech. 
Specifically, using the Phase 1 data, we compute three new, higher threshold values that target an average 3\% truncation rate for participants with different levels of A\&H severities: a \textit{mild} threshold based on the 25 Phase 1 participants with mild ratings, a \textit{moderate} threshold based on the 18 moderate participants, and a \textit{severe} threshold based on the 7 severe participants. By definition, increasing the threshold will always reduce or at worst maintain the same truncation rate as a lower threshold, at the cost of incurring a longer delay before the system responds. We evaluate these thresholds on the Phase 2 participant data.

As shown in Figure~\ref{fig:boxplots_endpointer_and_wer}(a), the new thresholds substantially reduce the truncation rate for PWS compared to the baseline. Even the smallest threshold increase, ``mild'', reduces the truncation rate to a per-participant average of 4.8\% ($SD$=6.4, $Median$=1.5, $IQR$=6.5), while the moderate threshold achieves our goal of under 3\% on average ($M$=2.5\%, $SD$=3.6, $Median$=0.8, $IQR$=3.1). 
Table~\ref{tab:endpointer} shows these results averaged across all Phase 2 utterances 
and includes metrics to capture how much delay occurs after the user finishes speaking. 
Here, $P50$ and $P95$ refer to 
the $\mathrm{50^{th}}$ (median) and $\mathrm{95^{th}}$ percentile delay between when a speech utterance ends and when the system stops listening. 
This analysis shows that the improvements in truncation rate come with a modest median delay of an additional 1.2 seconds over the baseline in the mild case and a 1.7 second increase in the moderate case. While the severe threshold is successful for 99.2\% of utterances, it causes a median delay of over 3 seconds. We return to these tradeoffs in the Discussion.

\begin{table*}[t!]
\begin{tabular}{|l|c|cc|}
\hline
\textbf{Endpointer Tuning} & \textbf{Truncation Rate} (\%) & \textbf{P50 Delay (ms)} & \textbf{P95 Delay (ms)} \\
\hline
General Population &  23.8                 & 450          & 1,270         \\
Stutter (mild) & 4.9                  & 1,670         & 2,750         \\
Stutter (moderate)  &2.5                  & 2,190         & 3,170         \\
Stutter (severe) &  0.8                  & 3,100         & 3,840        \\
\hline
\end{tabular}
\caption{Endpointer truncation rates and delays in milliseconds on all utterances in the Phase 2 dataset (n=5370 utterances) with threshold values tuned for different stuttering severities. There is a substantial decrease in truncations when moving from the endpointer tuned on the general population to one tuned on PWS who have mild stutter, then further reductions with tuning on moderate and severe; these reductions are consistent for every participant. P50 refers to the
$\mathrm{50^{th}}$ percentile (i.e., median) delay and P95 is the
$\mathrm{95^{th}}$ percentile delay. 
}
\label{tab:endpointer}
\end{table*}

\subsubsection{Baseline \& Improved ASR Model Performance} \label{sec:asr_baselines}
Next, we evaluated baseline ASR performance and compared the results to an approach that tunes the ASR decoder using dysfluent speech. For this analysis, we use the Apple Speech framework~\cite{SpeechFramework} and report on results from an ASR model trained on voice assistant tasks and speech from the general population. For completeness, we also examined baseline performance with a model trained on dictation tasks; the pattern of results was similar and thus results are omitted for clarity.

Our primary evaluation metric is Word Error Rate (WER), as widely used within the speech recognition community. 
\textit{Word Error Rate} is computed by counting the number of substitutions, insertions, and deletions in the transcribed text, and dividing by the total number of intended words.
For example, the intended phrase ``Add apples to my grocery list'' may be spoken ``A-(dd) a-(dd) add apples to my grocery list'' and be recognized as ``had had balls to my grocery list''. With two insertions and one misrecognized words, the WER is 50.0\%.\footnote{We report WER by first calculating each participant's individual WER, then computing averages across participants. This approach is common in the HCI community because it provides an understanding of the distribution of how individual users may experience the system. In contrast, the speech community often reports WER across an entire dataset without grouping by participant. Thus, for completeness and in line with the speech community's reporting, the baseline WER for all 91 participants' data combined is 19.9\%; for only Phase 2, all data combined yields baseline WER of 25.4\%. These numbers end up being close to the averaged WERs presumably due to similar number of phrases that each participant recorded.}
For some experiments, we also look at Thresholded WER, as used in work by Project Euphonia~\cite{green2021automatic,tobin2022personalized}, to assess VA performance for people with speech disabilities. This is computed as the percentage of utterances, per person, with WER below 10\% or 15\% (as specified). These values have been suggested as potential minimums for VAs to be useful depending on domain.
Lastly, we look at
\textit{Intent Error Rate} (IER), which captures whether the VA carries out the correct action in response to an utterance. To compute Intent Error Rate, we run the ASR output on the NLU model from Chen \etal~ \cite{chen2019active} which also relied on models used by the Apple Speech Framework.
\footnote{
While intent recognition is not the final response of the VA, it approximates what action would be performed and can be treated as an upper bound on command accuracy.
Recall that intents are used by VAs and encode the grammar of the VA command.
For example, the query ``What day is Christmas?'' may have the intent structure \lstinline{clock[clockDate + clockNoun(date) + clockVerb(finddate)]} where \textit{date} is Christmas and \textit{finddate} is an action. 
The VA would then execute and return the date: December $\mathrm{25^{th}}$. 
}


\paragraph{Baseline ASR}
Table~\ref{tab:baselineWERbySeverity} shows baseline ASR performance. Across participants in both phases, the average baseline WER is 19.8\%, 
which is much higher than the $\sim$5\% reported for consumer VA systems \cite{Xiong2016HumanParityASR, Stolcke2017HumanMachineComparison, Xiong2018MicrosoftASR}.   
The WER distribution is also highly skewed, with many participants having low WERs but also a long tail of participants with much higher WERs.
For people with mild A\&H severity, the average WER was 4.8\%, which is similar to what is expected for people who do not stutter, whereas moderate and severe had average WERs of 13.6\% and 49.2\%, respectively. Moreover, 84.0\% of participants with mild severity had a WER of less than 10\% (i.e., Thresholded WER) and the average Intent Error Rate for this group was 4.9\%, suggesting that many people with mild A\&H severity ratings would likely be able to use off-the-shelf VA systems, which echos the VA usage reported in survey presented in Figure ~\ref{fig:usage_frequency_by_severity_future_tech}(left).For moderate and severity grades, Intent Error Rates are 7.3\% and 18.4\% respectively. 
Overall, this analysis both confirms the ASR accuracy difficulties described in the survey, as well as reflects the varied experiences that survey participants reported in how well VAs understood them (Figure~\ref{fig:freqUseAndPerceivedAccuracyVAandDictation}). 
 

\begin{table*}[t!]
\begin{tabular}{|l|cccc|cc|cccc|}
\hline
& \multicolumn{4}{c}{\textbf{Word Error Rate (WER) (\%) }} & \multicolumn{2}{|c|}{\textbf{Thresholded WER (\%) }} & \multicolumn{4}{|c|}{\textbf{Intent Error Rate (IER) (\%) }} \\
\textbf{Participant Subset} & \textit{Mean} & \textit{SD} & \textit{Median} & \textit{IQR}  & \textit{WER<10} & \textit{WER<15} & \textit{Mean} & \textit{SD} & \textit{Median} & \textit{IQR} \\
\hline
Mild (n=25) & 4.8 & 6.7 & 2.4 & 1.8 & 84.0 & 88.0 & 4.9 & 4.2 & 6.1 & 6.9 \\
Moderate (n=44) & 13.6 & 16.7 & 8.5 & 12.2 & 63.6 & 75.0 & 7.3 & 6.8 & 5.0 & 9.0 \\
Severe (n=22) & 49.2 & 49.2 & 47.0 & 61.1 & 27.3 & 31.8 & 18.4 & 15.6 & 19.8 & 28.4 \\
\hline
All (n=91) & 19.8 & 31.7 & 7.1 & 20.2 & 60.4 & 69.2 & 9.3 & 10.5 & 6.1 & 9.9 \\
Phase 2 (eval set; n=41) & 25.4 & 28.7 & 10.8 & 33.8 & 48.8 & 61.0 & 10.4 & 10.8 & 5.3 & 13.4 \\
\hline
\end{tabular}
\caption{Baseline ASR model performance, broken down by severity level for all 91 participants, and overall for all 91 and for only Phase 2 participants. Participants with a mild A\&H rating had WERs on par with the general population, whereas more moderate and severe dysfluency speech greatly affects WER. Because we used Phase 1 for tuning our models, much of the analysis focuses on performance for the Phase 2 participants, who all had moderate or severe A\&H severity ratings.
}
\label{tab:baselineWERbySeverity}
\end{table*}





To understand how different types of dysfluencies affect WER, we examined the Phase 2 data, which includes detailed dysfluency annotations (see Section~\ref{sec:speech_data} for rates of each dysfluency type). Note that Phase 2 only included participants with moderate and severe A\&H ratings, so on average has somewhat higher WER (25.4\%) than the full set of participants (bottom row of Table~\ref{tab:baselineWERbySeverity}). 
For these Phase 2 participants, we found high Spearman's rank correlations\textsuperscript{\ref{spearman_footnote}} between WER and part-word repetitions ($r(39)=.85$, $p<.001$) and between WER and whole word repetitions ($r(39)=.60$, $p<.001$). The correlations were not significant for prolongations ($r(39)=.3$, $p=.056$), blocks ($r(39)=.21$, $p=.181$) or interjections ($r(39)=.15$, $p=.347$). 
This indicates that part-word and whole word repetitions tend to increase word error rates, and that blocks, prolongations and interjections
have less of an impact even if they are frequent. 


Among all errors the ASR system made in Phase 2, 80.9\% were word insertions, 17.5\% substitutions, and only 1.6\% deletions. The high rate of word insertions has a strong correlation with part-word repetitions ($r(39)=.73$, $p<.001$) followed by whole word repetitions ($r(39)=.60$, $p<.001$).
This echoes survey reports that part-word or whole word repetition can lead to misrecognized words being inserted in their transcription. 
A trained speech-language pathologist characterized insertion errors from part-word repetitions and found that in many cases insertions come from individual syllables being recognized as whole words (e.g., the first syllable in ``become'', vocalized as ``be-(come) be-(come) become''). 
In contrast, a sound repetition on /b/ in ``become'' is less likely to lead to word insertions. 
Furthermore, some people demonstrated part-word repetitions between syllables in multi-syllabic words, such as the word ``vocabulary'' may result in spurious insertions, such as ``vocab cab Cavaleri.''

\paragraph{ASR Decoder Tuning (Intervention \#2)}
Consumer ASR systems are commonly trained on thousands of hours of speech from the general public; an amount far larger than can likely be obtained from PWS.
However, an initial investigation by Mitra \etal~\cite{mitra2021} on 18 PWS has shown it may be possible to tune a small number of ASR decoder parameters to improve performance for PWS. 
Here, we validate this approach on our larger dataset and find even greater gains when tuning on our 50 participant Phase 1 subset.
While we defer to that paper for details,  
in brief the approach increases the importance of the language model relative to the acoustic component in the decoder and increases the penalty for word insertions.
These changes reduce the likelihood of predicting extraneous low-confidence words often caused by part-word repetitions and bias the model towards more likely voice assistant queries. 
We used Phase 1 data to tune these parameters and report results on Phase 2. 

The average WER for Phase 2 participants jumps from the baseline of 25.4\% to 12.4\% ($SD$=12.3, $median$=6.1, $IQR$=13.1) with the tuned decoder, which is a relative improvement of 51.2\%. A Wilcoxon signed rank test shows that this improvement is statistically significant. 
See Table~\ref{tab:WERbyExperimentAndbyDysfluencyType} for more metrics and to understand how WER improves as a function of dysfluency types. For example, the WER lowers (improves) in 43.2\% of utterances with part-word repetitions and increases (worsens) 3.9\% of the time. For ASR tuning, WER improves the most in utterances that have whole word repetitions, part-word repetitions, and interjections and least for those with prolongations or blocks.

\begin{table*}[]
\begin{tabular}{|l|cc|cc|cc|}
\hline
& \multicolumn{2}{c}{\textbf{ASR Decoder Tuning}} & \multicolumn{2}{|c|}{\textbf{Dysfluency Refinement}} & \multicolumn{2}{|c|}{\textbf{Combined}}  \\
\textbf{Dyfluency Type} & \textit{Improved} & \textit{Regressed} & \textit{Improved} & \textit{Regressed}  & \textit{Improved} & \textit{Regressed} \\
\hline
Part-word Repetition & 43.2\% & 3.9\% & 25.2\% & 0.0\% & 47.6\% & 3.5\%\\
Whole Word Repetition & 64.0\% & 3.0\% & 64.7\% & 0.3\% & 81.1\% & 2.0\% \\
Prolongation  & 30.9\% & 5.1\% & 12.9\% & 5.0\% & 32.8\% & 0.1\% \\
Block & 31.4\% & 5.2\% & 17.7\% & 0.1\% & 34.6\% & 5.0\% \\
Interjection & 53.3\% & 5.0\% & 30.9\% & 0.0\% & 57.2\% & 4.6\%\\
\hline
\end{tabular}
\caption{Percentage of utterances labeled with a given dysfluency type that result in \textit{improved} WER or \textit{regressed} (worsened) WER with: ASR decoder tuning, dysfluency refinement, or the combination of the two. Generally, the improvements come with few regressions.}
\label{tab:WERbyExperimentAndbyDysfluencyType}
\end{table*}

\subsubsection{Dysfluency Refinement (Intervention \#3)}
According to our survey, PWS are often displeased when seeing repeated words, phrases and filler words in their dictated notes and texts. 
To address this issue, we refine the transcribed text using two strategies.
First, we look at filler words such as ``um,'' ``eh,'' ``ah,'' ``uh'' and minor variations.
Many of these filters are not explicitly defined in the language model (i.e., by design ``eh'' is never predicted), however, in practice, short fillers are frequently transcribed as the word ``oh.'' As part of our approach, we remove ``oh'' from predictions, unless ``oh'' is used to represent the number zero.
We considered others such as ``like'' and ``you know,'' which are common in conversational speech, but they did not appear as fillers in our dataset, likely because the utterances tend to be short and more defined than free-form speech. 
Second, we remove repeated words and phrases. This is more challenging because words may naturally be repeated (e.g., ``We had had many discussions'').
In our refinement approach, we take all adjacent repeated words or phrases in a transcript and compute the statistical likelihood that they would appear consecutively in text using an n-gram language model that is similar to~\cite{Heafield2013LM}. 
If the probability is below a threshold,\footnote{The threshold $\tau$ was chosen based on early experiments on Phase 1 and non PWS-centered datasets.} i.e., $\mathrm{P}(\textsc{substring}_1, \textsc{substring}_2) < \tau$, then we remove the duplicate.
Note that these two strategies---interjection removal and repetition removal---can be applied to any ASR model output, thus we evaluated our dysfluency refinement approach in combination with both ASR models from Section~\ref{sec:asr_baselines}: the baseline model and the model with the tuned decoder.

As shown in Figure ~\ref{fig:boxplots_endpointer_and_wer}(b), dysfluency refinement reduces WER for the Phase 2 data compared to both the baseline and tuned ASR models. The average WER goes from a baseline of 25.4\% to 18.1\% ($SD$=19.2, $median$=6.1, $IQR$=22.8) after dysfluency refinement---a 28.7\% relative improvement on average. This improvement is significant with a Wilcoxon signed rank test ($W$=839, $Z$=5.43, $p$<.001).
As shown in Table~\ref{tab:WERbyExperimentAndbyDysfluencyType}, across the entire Phase 2 dataset, 64.7\% of utterances that contain whole word repetitions see a WER improvement and only 0.3\% regress. For those with interjections, WER improves in 30.9\% and regresses in none of the utterances. 

Applying dysfluency refinement to the output from the tuned ASR model further reduces WER, from on average 12.5\% with the tuned model alone to 9.9\% ($SD$=8.8, $median$=5.3, $IQR$=10.3) for the tuned model plus dysfluency refinement. Compared to the baseline ASR model, this tuned model plus dysfluency refinement combination is an average 61.2\% improvement across participants. 
Wilcoxon signed rank tests show that these improvements are statistically significant both when appended to the baseline ASR model output ($W$=741, $Z$=5.37, $p$<.001) or appended to the tuned ASR model output ($W$=595, $Z$=5.08, $p$<.001).
The percentage of participants with WER$<$10\% increases from 48.8\% to 65.9\%
and the average Intent Error Rate improved from 10.4\% ($SD$=10.8, $median$=5.3, $IQR$=13.4) to 5.4\% ($SD$=5.8, $median$=3.1, $IQR$=7.2). 
Such improvements may enable a VA to be usable for many of our participants when the baseline is not; for example, P2-21's WER improves from 40.4\% (baseline) to 13.0\% (tuned ASR + dysfluency refinements) and IER improves from 19.1\% (baseline) to 4.6\%.


\subsubsection{ASR Results Over Time}\label{sec:asr_over_time}
The above analyses assess the effectiveness of three changes to an existing speech recognition system that require little data from PWS to implement. 
At the same time, there has been substantial progress in general speech recognition models in recent years, with some accounts claiming WERs of 5\% or less using consumer voice assistants~\cite{Xiong2018MicrosoftASR}. Theoretically, these general improvements could also result in improvements for PWS. 

\begin{figure}[t!]
    \centering
    \includegraphics[width=0.4 \paperwidth]{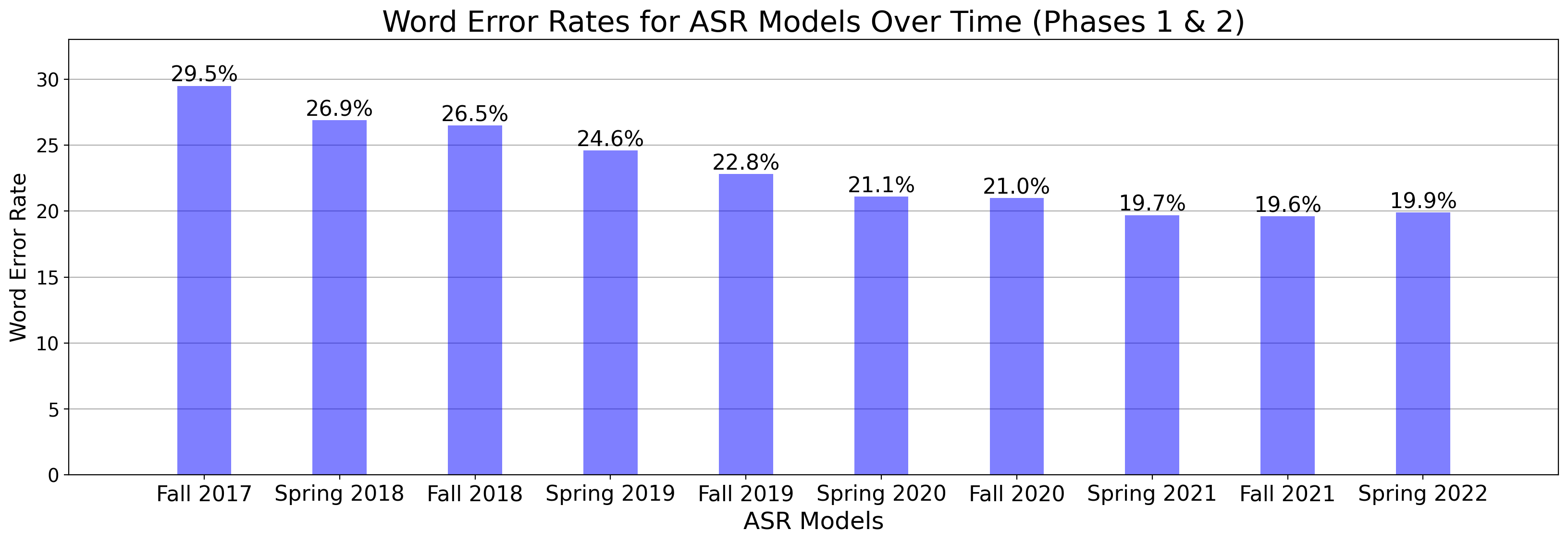}
    \caption{
    Performance of the speech recognition system as trained at different points between 2017 and 2022. There is a significant improvement in part- and whole word repetitions. Note that absolute WERs and their improvements may not be representative of performance on other sub-populations in the general public, which has been documented to be as low as 5\% WER (e.g., \cite{Xiong2018MicrosoftASR}). 
    }
    \label{fig:ASROverTime}
    \Description{This bar chart shows ASR performance over 10 points in time, from Fall 2017 to Spring 2022. Results gradually decrease from 29.5\% to 19.9\%, with a low of 19.6\% in the second most recent time point.
    }
\end{figure}

To understand how these changes over time impact performance on dysfluent speech,
we use the Apple Speech framework to run all Phase 1 and 2 data through archived ASR models that had been publicly available between Fall 2017 and Spring 2022; these experiments were conducted in Summer 2022.
Figure~\ref{fig:ASROverTime} shows WER across all utterances in the dataset every 6 months across the five years. The utterance-weighted WER was 29.5\% for the Fall 2017 model, fell consistently for the following eight time periods, and ended at 19.9\%. This is a 32.5\% relative reduction in WER from the start to the end of the 5-year time span. 
Differences at each time point may be attributed to what data was used in training, the convolutional architecture, and/or the language model. 

For Phase 2 utterances, we further examined changes in how specific types of dysfluencies manifest in WER performance between the Fall 2017 to Spring 2022 models. For part-word repetitions, the WER improved 43.5\% of utterances (worse 12.2\%), those with whole word repetitions 46.8\% (worse 14.5\%), prolongations 36.9\% (worse 10.3\%), blocks 36.0\% (worse 9.4\%), and interjections 65.8\% (worse 10.4\%).
The improvements with part-word repetitions are especially interesting, because these errors are more challenging to correct using our strategies. 
See the Discussion (Section~\ref{sec:discussion}) for further implications of these findings. 

\subsection{Summary}

This performance analysis mirrors the primary concerns raised by survey participants and identifies promising directions for improving speech recognition performance for PWS. 
With baseline production-grade models, data from PWS resulted in high truncation rates (23.8\%), often due to blocks, as well as high WER rates (19.8\%), which were dominated by insertion errors. 
Correlating these errors with dysfluency annotations highlighted how different stuttering characteristics can impact different types of errors. 
Fortunately, the three interventions investigated offer improvements for PWS, and can be implemented with limited data and changes to an ASR system. By tuning the endpointer detection threshold on data from PWS, truncation rates on Phase 2 data had a relative average drop of 79.1\% (``mild'') with minor additional system delay of 1.2 seconds.
Both ASR decoder tuning and dysfluency refinement improve WER compared to the baseline ASR model transcriptions, and combining the two offers a further reduction in WER (relative drop of 61.2\% compared to the baseline ASR model alone for Phase 2 data).




\section{Discussion}
\label{sec:discussion}

Our findings demonstrate the contrasting experiences that PWS encounter with speech technologies, and an overall strong interest in improved speech input accessibility from most survey participants. Off-the-shelf VAs and dictation work reasonably well for many people with mild stuttering severity. The survey shows that many PWS use speech recognition regularly (32.2\% daily), albeit at a lower rate compared to the general population ~\cite{npr_smart_audio}. Moreover, the performance analysis shows that most (84.0\%) participants with mild stuttering severity had word error rates of less than 10\%, a suggested threshold below which VAs can become usable~\cite{green2021automatic}. 

At the same time, participants with more moderate to severe speech dysfluencies often encountered serious technical barriers to adoption. Being cut off too early (with VAs) and ASR accuracy (both VAs and dictation) were primary technical concerns reflected in both the survey and baseline speech analysis: for the Phase 2 participants, truncation rates were substantially higher than the targeted general population rate (23.8\% vs. 3\%) and average word error rate was 25.4\%. We investigated three promising interventions (two new and one previously introduced) to address these issues: (1) tuning the endpointer threshold on moderate dysfluent speech reduces truncations for PWS to below the target of 3\%, (2) tuning the ASR decoder (approach from \cite{mitra2021}) significantly reduces word error rate, and (3) applying dysfluency refinements to the ASR output further significantly reduces word error rate regardless of whether we are using the baseline ASR model or the tuned ASR model. Also of importance, however, our survey highlights that regardless of how well speech technology works, some PWS will still not be interested in it due to perceptions of utility, the physical effort required to speak, and not wanting to use speech input around other people.

Overall, this work is the first to quantify user experiences for PWS, connect those experiences to performance across a range of stuttering severities, and propose and show the impact of multiple technical improvements to support PWS in using speech technology.
We reflect on these findings and potential future research in technical improvements for dysfluent speech and in the design and user experience of such systems. We also cover limitations of the work.

\subsection{Technical Improvements for Dysfluent Speech} 

Our approach has been to identify technical improvements that integrate with rather than replace existing systems, by offering relatively small adaptations to the speech recognition pipeline---changing an endpointer threshold, tuning the ASR's decoder model, and inserting a post-processing step to refine dysfluencies. Moreover, the techniques require very little data compared to what would be needed to train endpointer and ASR models from scratch, and are thus in contrast to many recent papers that replace existing systems with large ASR models trained on dysfluent speech (e.g., \cite{shonibare2022enhancing,alharbi2018lightly,Mendelev2020}) or personalized ASR models that require large amounts of data from a specific person (e.g., \cite{green2021automatic} which requires 15 - 120 minutes of speech). 



As shown in Section \ref{sec:asr_over_time}, general ASR improvements do improve performance for PWS, but a substantial gap remains between current and desired performance. While this gap may continue to narrow over time, concerns unique to PWS will likely persist. 
For example, multiple respondents, unprompted, highlighted how word repetitions and interjections in their transcriptions lead to ``errors'' and can result in increased stress or emotional response. 
General ASR modeling innovations typically transcribe verbatim, thus including these unintended words.
As such, human-centered ASR approaches like ours are likely necessary for improving the overall speech recognition experience for PWS. 

The techniques we investigated, while promising, also come with tradeoffs that benefit some users but may degrade performance for others. 
For example, a user who frequently experiences blocks may be happy to increase a VA system's delay in responding if it means that they will be able to use the system more effectively, whereas someone who rarely blocks may be annoyed by the slower response time.
While ideally a single ASR system could apply to all users, one short-term approach is to offer individuals an option to opt-in to specific settings that may benefit them. 
In this case, allowing users to choose between a small set of endpointer options may be useful. 

For ASR decoder tuning, tradeoffs may exist between improved recognition of dysfluent words (e.g., due to part-word repetitions) and a possibly increased error rate for rarer words due to the higher language model weight.
And, for dysfluency refinement, the system could inadvertently eliminate some words that appear to be dysfluencies but were intentional on the user's part (e.g., removing the duplicate ``so'' in ``I am so so happy'').
In the short-term, ASR decorder tuning and dysfluency refinement could also be enabled as opt-in features, though future work should investigate how to seamlessly integrate them into general ASR systems. 
For example, an improved dysfluency refinement approach could learn to distinguish dysfluent from intended repetitions, handle partial word repetitions in addition to whole word repetitions and interjections, and even learn filler word patterns unique to a given person. 

We focused on endpointing and ASR solutions, but two additional technical concerns that arose with less frequency in the survey data are wake words (e.g., ``Alexa'', ``Hey Siri'') and automatic punctuation. Offering multiple wake word alternatives, custom wake words, or flexibility in the pause length between words so as to accommodate blocks could improve accessibility for these users. 
For automatic punctuation, multiple respondents mentioned punctuation being inserted incorrectly during pauses in speech. 
Effort should be made to understand how blocks and pauses impact current punctuation models and if there are ways to avoid incorrectly inserted punctuation.

\subsection{Additional Design and User Experience Considerations}








A key next step will be to validate our speech experiments by incorporating the three technical improvements into a fully interactive system to evaluate with PWS. Such an evaluation will allow us to understand the extent to which the offline experiments here translate to impacts on user experience and satisfaction with speech input---fully closing the loop with the subjective experiences reported in the survey. Understanding how users feel about the tradeoffs described above (e.g., truncation rate vs. response delay) and what dysfluencies they want to have refined in their transcribed speech will be important. Preferences may vary greatly across individuals, for example, with some not wanting the system to ``fix'' their transcribed speech.

In addition to the above model-side improvements for PWS, there is potential for some challenges to be addressed through more interactive approaches. One possibility is for users to provide feedback directly to the system, training it to better support their own stuttering patterns and to accommodate their desire for different types of refinements. For dictation and screen-based VAs, this type of feedback could occur by displaying the transcribed speech and allowing the user to manipulate it through direct touch to indicate error words that should be removed or corrected \cite{dictation_user_expectations}. 
Perhaps the first time a pattern is encountered, such as a suspected full or partial repetition of a specific word or a transcribed interjection from a list of common interjections (e.g., ``um'', ``like''), the system could prompt the user to indicate that it should hide or repair future similar dysfluencies from transcripts. 

Correction feedback on screen-free voice assistant devices is more difficult, because the user may not see what words were predicted; they may only receive notification of what VA action was performed.  
One possibility would be to learn how speakers rephrase their command when the voice assistant suggests a wrong action, which could be useful if an individual intentionally chooses new words that they are less likely to stutter on. 
Some current systems also allow users to review the recognized text on another device (e.g., Alexa's smartphone app), and could potentially be extended to allow users to correct the transcription. 
Any of these forms of correction could ultimately be incorporated into the system to personalize speech recognition.

The design of any speech interface for PWS will need to also account for non-technical contextual factors around use. As demonstrated in survey responses, some PWS may find their ability or desire to use speech technology will differ depending on whether other people are around (some respondents did not want others to hear them speak). Further, some participants reported that their stuttering severity varied based on their history with the device, for example, increasing following a negative interaction with the device. Both of these scenarios emphasize the importance of supporting PWS in seamlessly transitioning from speech to text-based input as needed, and ensuring that speech is never the only means of accomplishing a task.

Emphasizing the utility of combining multiple modalities of input for PWS, most survey participants (84.7\%) used VAs on a smartphone, and some responses pointed to the accessibility benefits of doing so. Speech affordances vary greatly on smartphones versus smart speakers, with smartphones offering alternative invocation mechanisms such as holding a physical button to speak or pushing a virtual button on-screen.
In addition, holding a button (e.g., on iOS)
can disable the endpointer, meaning a user will not be cut off mid-utterance. 
These features are also available on some smartwatches, but not at-a-distance devices like smart speakers. 
Future work should investigate alternatives that improve these experiences on screen-free devices.

\subsection{Beyond Fluency}
While the primary focus of this paper is on stuttering, the interventions may improve speech recognition for people with other types of communication disabilities. 
For example, endpointer improvements may improve performance for people with hypokinetic dysarthria that experience unpredictable silences (e.g., Parkinson's Disease) or hyperkinetic dysarthria that experience intermittent aphonia or voice stoppages in their speech (e.g., Tourette syndrome) \cite{duffy2019motor}.
These improvements may also affect individuals with expressive aphasia who take longer to come up with a target word \cite{evans2020much}.
Dysfluency refinement could improve WER for individuals with spasmodic dysphonia and hypokinetic dysarthia as they may produce part- or whole word repetitions~\cite{goberman2010characteristics,juste2018frequency}.
People with spasmodic dysphonia may present with an increase in interjections \cite{cannito1997disfluency}, which could be removed with Dysfluency Refinement or ASR Tuning. 

Other speech, voice, or language differences not addressed by this work include:
breathy breaks mid-word that can cause word insertions and substitutions (e.g., in cases of spasmodic dysphonia \cite{murry2014spasmodic}), 
distorted vowels which could result in word substitutions  (e.g., in cases of ataxic and hyperkinetic dysarthria \cite{duffy2019motor}), 
pitch breaks (e.g., in cases of flaccid dysarthria and some voice disorders \cite{duffy2019motor}),  
atypical rhythm and stress (e.g., in cases of flaccid and hypokinetic dysarthria \cite{darley1969differential}),
speech characteristics that deteriorate or change over time (e.g., in cases of Amytrophic Lateral Sclerosis (ALS), some voice disorders, and some dysathrias), and 
vowel neutralization and differences in resonance (e.g, in cases of speech from Deaf/Hard of Hearing individuals \cite{subtelny1992spectral}).

\subsection{Limitations and Potential Biases}
\label{discussion_limitations}

Speech characteristics can vary with time and with settings, as noted above and in the literature (e.g.,~\cite{tichenor2021variability,bleakley2022exploring}), so our collected audio data does not provide a full view of speech characteristics for each participant. 
Recordings were done at an individual's convenience, while by themselves, which likely impacted the prevalence and severity of dysfluencies.
Some users may have had more dysfluencies because they knew they were being recorded, while others may have had less because they could do it when they felt most comfortable. 
One improvement on our methodology would be to ask participants to record commands at different times of the day, in different environments, and perhaps different situations where their dysfluency patterns might be different. We also focused solely on speech accessibility and did not ask participants if they had other accessibility needs for computing devices---which could impact the effectiveness of some solutions like pushing a button to invoke a voice assistant.

The fine-grained dysfluency annotations enabled detailed analysis of how rates vary per-person, however, there were two types of annotator errors that we estimate to have impacted a small percent of clips. 
We found that some dysfluency types conflated for one-another. For example, some sound or part-word repetitions (e.g., ``s-s-snake'') were conflated with prolongations (e.g., ``ssssnake'') and vice versa. 
This means that there may be discrepancies between our reported dysfluency correlations and the actual correlations. 
Second, with the transcript annotations, people were asked to label both the articulated and the ``intended'' speech. Ideally, the intended speech would have been verified by participants; instead it was labeled with an annotator's best guess based on context. 
In a small percentage of clips we found likely discrepancies and in many cases improved with an annotation refinement pass. 

Quality of ASR results varies significantly as a function of uniqueness; common VA queries or dictation phrases may be more likely to be accurately recognized and rare words or phrases are less likely. 
This is even more the case with our ASR tuning approach, which biases the language modeling priors over the acoustic information. 
Given that the majority of the audio recording prompts were VA-related---albeit with proper nouns such as musician names---our WER may be lower than what one might expect in the real-world, especially for dictation which is more open ended. As emphasized above, a key step for future work will be to evaluate the improvements proposed here with PWS using a fully interactive system. 

The findings in Section \ref{sec:asr_results} are derived using an ASR system with a hybrid architecture, specifically with the Apple Speech Framework. 
ASR systems trained with different data or architectures could exhibit different model biases, as seen in our analysis in Section \ref{sec:asr_over_time}, where each model exhibited different performances and biases toward certain dysfluency types. Nonetheless, our general findings of poor WER performance for stuttered speech is congruent to similar WER analysis conducted on other ASR systems using different architectures \cite{shonibare2022enhancing, AlharbiWOCCI17,alharbi2018lightly}.

\section{Conclusion}

In this paper, we described survey findings from 61 people who stutter and performed speech recognition experiments on a 91-person collection. We found that some PWS already use speech recognition systems regularly, some state that they would never be interested, but regardless of existing performance over half of respondents would use speech technology more frequently if they were more accurate. 
While our speech recognition results indicated worse baseline performance relative to the general public, especially for people with higher rates of dysfluencies, our model tuning and dysfluency refinement solutions can drastically reduce endpointing and ASR error rates, and ultimately enable ``good enough'' (WER$<$15) performance for most users. 

\section*{Acknowledgements}
Thanks to Shrinath Thelapurath, Zeinab Liaghat, Hannah Gillis Coleman, Adriana Hilliard, Chris Maury, Carolyn Feng, Gautam Varma Mantena, and Leo Liu for various discussions and study related tasks.


% ...

\bibliographystyle{ACM-Reference-Format}
\bibliography{main.bib}


\begin{thebibliography}{72}


\ifx \showCODEN    \undefined \def \showCODEN     #1{\unskip}     \fi
\ifx \showDOI      \undefined \def \showDOI       #1{#1}\fi
\ifx \showISBNx    \undefined \def \showISBNx     #1{\unskip}     \fi
\ifx \showISBNxiii \undefined \def \showISBNxiii  #1{\unskip}     \fi
\ifx \showISSN     \undefined \def \showISSN      #1{\unskip}     \fi
\ifx \showLCCN     \undefined \def \showLCCN      #1{\unskip}     \fi
\ifx \shownote     \undefined \def \shownote      #1{#1}          \fi
\ifx \showarticletitle \undefined \def \showarticletitle #1{#1}   \fi
\ifx \showURL      \undefined \def \showURL       {\relax}        \fi
\providecommand\bibfield[2]{#2}
\providecommand\bibinfo[2]{#2}
\providecommand\natexlab[1]{#1}
\providecommand\showeprint[2][]{arXiv:#2}

\bibitem[Abdolrahmani et~al\mbox{.}(2018)]%
        {abdolrahmani2018siri}
\bibfield{author}{\bibinfo{person}{Ali Abdolrahmani}, \bibinfo{person}{Ravi
  Kuber}, {and} \bibinfo{person}{Stacy~M Branham}.}
  \bibinfo{year}{2018}\natexlab{}.
\newblock \showarticletitle{``Siri Talks at You'' An Empirical Investigation of
  Voice-Activated Personal Assistant (VAPA) Usage by Individuals Who Are
  Blind}. In \bibinfo{booktitle}{\emph{Proceedings of the 20th International
  ACM SIGACCESS Conference on Computers and Accessibility}}.
  \bibinfo{pages}{249--258}.
\newblock


\bibitem[Alharbi et~al\mbox{.}(2018)]%
        {alharbi2018lightly}
\bibfield{author}{\bibinfo{person}{Sadeen Alharbi}, \bibinfo{person}{Madina
  Hasan}, \bibinfo{person}{Anthony~JH Simons}, \bibinfo{person}{Shelagh
  Brumfitt}, {and} \bibinfo{person}{Phil Green}.}
  \bibinfo{year}{2018}\natexlab{}.
\newblock \showarticletitle{A lightly supervised approach to detect stuttering
  in children's speech}. In \bibinfo{booktitle}{\emph{Interspeech 2018}}. ISCA,
  \bibinfo{pages}{3433--3437}.
\newblock


\bibitem[Alharbi et~al\mbox{.}(2017)]%
        {AlharbiWOCCI17}
\bibfield{author}{\bibinfo{person}{Sadeen Alharbi}, \bibinfo{person}{Anthony
  J.~H. Simons}, \bibinfo{person}{Shelagh Brumfitt}, {and}
  \bibinfo{person}{Phil~D. Green}.} \bibinfo{year}{2017}\natexlab{}.
\newblock \showarticletitle{Automatic recognition of children's read speech for
  stuttering application}.
\newblock \bibinfo{journal}{\emph{International Workshop on Child Computer
  Interaction}}  \bibinfo{volume}{WOCCI} (\bibinfo{year}{2017}).
\newblock


\bibitem[Ammari et~al\mbox{.}(2019)]%
        {AmmariCHI21}
\bibfield{author}{\bibinfo{person}{Tawfiq Ammari}, \bibinfo{person}{Jofish
  Kaye}, \bibinfo{person}{Janice~Y. Tsai}, {and} \bibinfo{person}{Frank
  Bentley}.} \bibinfo{year}{2019}\natexlab{}.
\newblock \showarticletitle{Music, Search, and IoT: How People (Really) Use
  Voice Assistants}.
\newblock \bibinfo{journal}{\emph{ACM Transactions on Computer-Human
  Interaction.}} \bibinfo{volume}{26}, \bibinfo{number}{3}, Article
  \bibinfo{articleno}{17} (\bibinfo{date}{apr} \bibinfo{year}{2019}),
  \bibinfo{numpages}{28}~pages.
\newblock
\showISSN{1073-0516}


\bibitem[Anantha et~al\mbox{.}(2020)]%
        {ranantha2020intent}
\bibfield{author}{\bibinfo{person}{Raviteja Anantha}, \bibinfo{person}{Srinivas
  Chappidi}, {and} \bibinfo{person}{Arash~W Dawoodi}.}
  \bibinfo{year}{2020}\natexlab{}.
\newblock \showarticletitle{Learning to Rank Intents in Voice Assistants}. In
  \bibinfo{booktitle}{\emph{Conversational Dialogue Systems for the Next
  Decade}}. \bibinfo{publisher}{Springer Singapore},
  \bibinfo{address}{Singapore}, \bibinfo{pages}{87--101}.
\newblock


\bibitem[Andrews and Harris(1964)]%
        {AndrewsHarris1964}
\bibfield{author}{\bibinfo{person}{Gavin Andrews} {and} \bibinfo{person}{Mary
  Harris}.} \bibinfo{year}{1964}\natexlab{}.
\newblock \showarticletitle{The syndrome of stuttering.}
\newblock \bibinfo{journal}{\emph{Spastics Society Medical Education}}
  (\bibinfo{year}{1964}).
\newblock


\bibitem[Apple(2022)]%
        {SpeechFramework}
\bibfield{author}{\bibinfo{person}{Apple}.} \bibinfo{year}{2022}\natexlab{}.
\newblock \bibinfo{title}{Speech Framework}.
\newblock
  \bibinfo{howpublished}{\url{https://developer.apple.com/documentation/speech}}.
\newblock


\bibitem[Arjun et~al\mbox{.}(2020)]%
        {Tripathi2020}
\bibfield{author}{\bibinfo{person}{KN Arjun}, \bibinfo{person}{S Karthik},
  \bibinfo{person}{D Kamalnath}, \bibinfo{person}{Pranavi Chanda}, {and}
  \bibinfo{person}{Shikha Tripathi}.} \bibinfo{year}{2020}\natexlab{}.
\newblock \showarticletitle{Automatic Correction of Stutter in Disfluent
  Speech}.
\newblock \bibinfo{journal}{\emph{CoCoNet}} (\bibinfo{year}{2020}).
\newblock


\bibitem[Ballati et~al\mbox{.}(2018)]%
        {ballati2018assessing}
\bibfield{author}{\bibinfo{person}{Fabio Ballati}, \bibinfo{person}{Fulvio
  Corno}, {and} \bibinfo{person}{Luigi De~Russis}.}
  \bibinfo{year}{2018}\natexlab{}.
\newblock \showarticletitle{Assessing virtual assistant capabilities with
  Italian dysarthric speech}. In \bibinfo{booktitle}{\emph{Proceedings of the
  20th International ACM SIGACCESS Conference on Computers and Accessibility}}.
  \bibinfo{pages}{93--101}.
\newblock


\bibitem[Basapur et~al\mbox{.}(2007)]%
        {dictation_user_expectations}
\bibfield{author}{\bibinfo{person}{Santosh Basapur}, \bibinfo{person}{Shuang
  Xu}, \bibinfo{person}{Mark Ahlenius}, {and} \bibinfo{person}{Young~Seok
  Lee}.} \bibinfo{year}{2007}\natexlab{}.
\newblock \showarticletitle{User Expectations from Dictation on Mobile
  Devices}. In \bibinfo{booktitle}{\emph{Human-Computer Interaction.
  Interaction Platforms and Techniques}},
  \bibfield{editor}{\bibinfo{person}{Julie~A. Jacko}} (Ed.).
  \bibinfo{publisher}{Springer Berlin Heidelberg}, \bibinfo{address}{Berlin,
  Heidelberg}, \bibinfo{pages}{217--225}.
\newblock
\showISBNx{978-3-540-73107-8}


\bibitem[Bayerl et~al\mbox{.}(2020)]%
        {BayerlTSD2020}
\bibfield{author}{\bibinfo{person}{Sebastian Bayerl}, \bibinfo{person}{Florian
  H{\"o}nig}, \bibinfo{person}{Joelle Reister}, {and}
  \bibinfo{person}{Korbinian Riedhammer}.} \bibinfo{year}{2020}\natexlab{}.
\newblock \showarticletitle{Towards Automated Assessment of Stuttering and
  Stuttering Therapy}. In \bibinfo{booktitle}{\emph{International Conference on
  Text, Speech, and Dialogue}}.
\newblock


\bibitem[Bayerl et~al\mbox{.}(2022)]%
        {KSoF}
\bibfield{author}{\bibinfo{person}{Sebastian~P Bayerl},
  \bibinfo{person}{Alexander~Wolff von Gudenberg}, \bibinfo{person}{Florian
  H{\"o}nig}, \bibinfo{person}{Elmar N{\"o}th}, {and}
  \bibinfo{person}{Korbinian Riedhammer}.} \bibinfo{year}{2022}\natexlab{}.
\newblock \showarticletitle{KSoF: The Kassel State of Fluency Dataset--A
  Therapy Centered Dataset of Stuttering}.
\newblock \bibinfo{journal}{\emph{arXiv preprint arXiv:2203.05383}}
  (\bibinfo{year}{2022}).
\newblock


\bibitem[Bleakley et~al\mbox{.}(2022)]%
        {bleakley2022exploring}
\bibfield{author}{\bibinfo{person}{Anna Bleakley}, \bibinfo{person}{Daniel
  Rough}, \bibinfo{person}{Abi Roper}, \bibinfo{person}{Stephen Lindsay},
  \bibinfo{person}{Martin Porcheron}, \bibinfo{person}{Minha Lee},
  \bibinfo{person}{Stuart Nicholson}, \bibinfo{person}{Benjamin Cowan}, {and}
  \bibinfo{person}{Leigh Clark}.} \bibinfo{year}{2022}\natexlab{}.
\newblock \showarticletitle{Exploring Smart Speaker User Experience for People
  Who Stammer}. In \bibinfo{booktitle}{\emph{Proceedings of the ACM SIGACCESS
  Conference on Computers \& Accessibility}}.
\newblock


\bibitem[Cannito et~al\mbox{.}(1997)]%
        {cannito1997disfluency}
\bibfield{author}{\bibinfo{person}{Michael~P Cannito},
  \bibinfo{person}{Annette~Renee Burch}, \bibinfo{person}{Christopher Watts},
  \bibinfo{person}{Patrick~W Rappold}, \bibinfo{person}{Stephen~B Hood}, {and}
  \bibinfo{person}{Kyla Sherrard}.} \bibinfo{year}{1997}\natexlab{}.
\newblock \showarticletitle{Disfluency in spasmodic dysphonia: A multivariate
  analysis}.
\newblock \bibinfo{journal}{\emph{Journal of Speech, Language, and Hearing
  Research}} \bibinfo{volume}{40}, \bibinfo{number}{3} (\bibinfo{year}{1997}),
  \bibinfo{pages}{627--641}.
\newblock


\bibitem[Chan et~al\mbox{.}(2015)]%
        {chan2015ListenAttendSpell}
\bibfield{author}{\bibinfo{person}{William Chan}, \bibinfo{person}{Navdeep
  Jaitly}, \bibinfo{person}{Quoc~V Le}, {and} \bibinfo{person}{Oriol Vinyals}.}
  \bibinfo{year}{2015}\natexlab{}.
\newblock \showarticletitle{Listen, attend and spell}.
\newblock \bibinfo{journal}{\emph{arXiv preprint arXiv:1508.01211}}
  (\bibinfo{year}{2015}).
\newblock


\bibitem[Chen et~al\mbox{.}(2019)]%
        {chen2019active}
\bibfield{author}{\bibinfo{person}{Xi~C Chen}, \bibinfo{person}{Adithya Sagar},
  \bibinfo{person}{Justine~T Kao}, \bibinfo{person}{Tony~Y Li},
  \bibinfo{person}{Christopher Klein}, \bibinfo{person}{Stephen Pulman},
  \bibinfo{person}{Ashish Garg}, {and} \bibinfo{person}{Jason~D Williams}.}
  \bibinfo{year}{2019}\natexlab{}.
\newblock \showarticletitle{Active learning for domain classification in a
  commercial spoken personal assistant}.
\newblock \bibinfo{journal}{\emph{Interspeech}} (\bibinfo{year}{2019}).
\newblock


\bibitem[Clark et~al\mbox{.}(2020)]%
        {clark2020speech}
\bibfield{author}{\bibinfo{person}{Leigh Clark}, \bibinfo{person}{Benjamin~R.
  Cowan}, \bibinfo{person}{Abi Roper}, \bibinfo{person}{Stephen Lindsay}, {and}
  \bibinfo{person}{Owen Sheers}.} \bibinfo{year}{2020}\natexlab{}.
\newblock \showarticletitle{Speech Diversity and Speech Interfaces: Considering
  an Inclusive Future through Stammering}. In
  \bibinfo{booktitle}{\emph{Proceedings of the 2nd Conference on Conversational
  User Interfaces}} (Bilbao, Spain) \emph{(\bibinfo{series}{CUI '20})}.
  \bibinfo{publisher}{Association for Computing Machinery},
  \bibinfo{address}{New York, NY, USA}, Article \bibinfo{articleno}{24},
  \bibinfo{numpages}{3}~pages.
\newblock
\showISBNx{9781450375443}
\urldef\tempurl%
\url{https://doi.org/10.1145/3405755.3406139}
\showDOI{\tempurl}


\bibitem[Corbett and Weber(2016)]%
        {corbett2016can}
\bibfield{author}{\bibinfo{person}{Eric Corbett} {and} \bibinfo{person}{Astrid
  Weber}.} \bibinfo{year}{2016}\natexlab{}.
\newblock \showarticletitle{What can I say? addressing user experience
  challenges of a mobile voice user interface for accessibility}. In
  \bibinfo{booktitle}{\emph{Proceedings of the 18th international conference on
  human-computer interaction with mobile devices and services}}.
  \bibinfo{pages}{72--82}.
\newblock


\bibitem[Corcoran(2018)]%
        {Slate}
\bibfield{author}{\bibinfo{person}{Moira Corcoran}.}
  \bibinfo{year}{2018}\natexlab{}.
\newblock \bibinfo{title}{When Alexa Can’t Understand You}.
\newblock \bibinfo{howpublished}{Slate (online)}.
\newblock


\bibitem[Craig et~al\mbox{.}(2002)]%
        {craig2002epidemiology}
\bibfield{author}{\bibinfo{person}{Ashley Craig}, \bibinfo{person}{Karen
  Hancock}, \bibinfo{person}{Yvonne Tran}, \bibinfo{person}{Magali Craig},
  {and} \bibinfo{person}{Karen Peters}.} \bibinfo{year}{2002}\natexlab{}.
\newblock \showarticletitle{Epidemiology of stuttering in the community across
  the entire life span}.
\newblock  (\bibinfo{year}{2002}).
\newblock


\bibitem[Darley et~al\mbox{.}(1969)]%
        {darley1969differential}
\bibfield{author}{\bibinfo{person}{Frederic~L Darley},
  \bibinfo{person}{Arnold~E Aronson}, {and} \bibinfo{person}{Joe~R Brown}.}
  \bibinfo{year}{1969}\natexlab{}.
\newblock \showarticletitle{Differential diagnostic patterns of dysarthria}.
\newblock \bibinfo{journal}{\emph{Journal of speech and hearing research}}
  \bibinfo{volume}{12}, \bibinfo{number}{2} (\bibinfo{year}{1969}),
  \bibinfo{pages}{246--269}.
\newblock


\bibitem[Das et~al\mbox{.}(2020)]%
        {DasNSRE2020}
\bibfield{author}{\bibinfo{person}{Arun Das}, \bibinfo{person}{Jeffrey Mock},
  \bibinfo{person}{Henry Chacon}, \bibinfo{person}{Farzan Irani},
  \bibinfo{person}{Edward Golob}, {and} \bibinfo{person}{Peyman Najafirad}.}
  \bibinfo{year}{2020}\natexlab{}.
\newblock \showarticletitle{Stuttering Speech Disfluency Prediction using
  Explainable Attribution Vectors of Facial Muscle Movements}. In
  \bibinfo{booktitle}{\emph{arXiv}}.
\newblock


\bibitem[Dash et~al\mbox{.}(2018)]%
        {Tripathi2018}
\bibfield{author}{\bibinfo{person}{Ankit Dash}, \bibinfo{person}{Nikhil
  Subramani}, \bibinfo{person}{Tejas Manjunath}, \bibinfo{person}{Vishruti
  Yaragarala}, {and} \bibinfo{person}{Shikha Tripathi}.}
  \bibinfo{year}{2018}\natexlab{}.
\newblock \showarticletitle{Speech Recognition and Correction of a Stuttered
  Speech}. In \bibinfo{booktitle}{\emph{2018 International Conference on
  Advances in Computing, Communications and Informatics ({ICACCI})}}.
  \bibinfo{pages}{1757--1760}.
\newblock


\bibitem[Deighton(2021)]%
        {WSJ}
\bibfield{author}{\bibinfo{person}{Katie Deighton}.}
  \bibinfo{year}{2021}\natexlab{}.
\newblock \bibinfo{title}{Tech Firms Train Voice Assistants to Understand
  Atypical Speech}.
\newblock \bibinfo{howpublished}{Wall Street Journal (online)}.
\newblock


\bibitem[Demarin et~al\mbox{.}(2015)]%
        {demarin2015impact}
\bibfield{author}{\bibinfo{person}{Iva Demarin}, \bibinfo{person}{Ljubica
  Leko}, \bibinfo{person}{Maja {\v{S}}krobo}, \bibinfo{person}{Helena Germano},
  \bibinfo{person}{Patr{\'\i}cia Macedo}, {and} \bibinfo{person}{Rui~Neves
  Madeira}.} \bibinfo{year}{2015}\natexlab{}.
\newblock \showarticletitle{The Impact of Stuttering; How Can a Mobile App
  Help?}. In \bibinfo{booktitle}{\emph{Proceedings of the 17th International
  ACM SIGACCESS Conference on Computers \& Accessibility}}.
  \bibinfo{pages}{399--400}.
\newblock


\bibitem[Duffy(2019)]%
        {duffy2019motor}
\bibfield{author}{\bibinfo{person}{Joseph~R Duffy}.}
  \bibinfo{year}{2019}\natexlab{}.
\newblock \bibinfo{booktitle}{\emph{Motor speech disorders e-book: Substrates,
  differential diagnosis, and management}}.
\newblock \bibinfo{publisher}{Elsevier Health Sciences}.
\newblock


\bibitem[Evans et~al\mbox{.}(2020)]%
        {evans2020much}
\bibfield{author}{\bibinfo{person}{William~S Evans}, \bibinfo{person}{William~D
  Hula}, \bibinfo{person}{Yina Quique}, {and} \bibinfo{person}{Jeffrey~J
  Starns}.} \bibinfo{year}{2020}\natexlab{}.
\newblock \showarticletitle{How much time do people with aphasia need to
  respond during picture naming? Estimating optimal response time cutoffs using
  a multinomial ex-Gaussian approach}.
\newblock \bibinfo{journal}{\emph{Journal of Speech, Language, and Hearing
  Research}} \bibinfo{volume}{63}, \bibinfo{number}{2} (\bibinfo{year}{2020}),
  \bibinfo{pages}{599--614}.
\newblock


\bibitem[Fok et~al\mbox{.}(2018)]%
        {fok2018towards}
\bibfield{author}{\bibinfo{person}{Raymond Fok}, \bibinfo{person}{Harmanpreet
  Kaur}, \bibinfo{person}{Skanda Palani}, \bibinfo{person}{Martez~E Mott},
  {and} \bibinfo{person}{Walter~S Lasecki}.} \bibinfo{year}{2018}\natexlab{}.
\newblock \showarticletitle{Towards more robust speech interactions for deaf
  and hard of hearing users}. In \bibinfo{booktitle}{\emph{Proceedings of the
  20th international ACM SIGACCESS conference on computers and accessibility}}.
  \bibinfo{pages}{57--67}.
\newblock


\bibitem[Ghai and Mueller(2021)]%
        {ghai2021fluent}
\bibfield{author}{\bibinfo{person}{Bhavya Ghai} {and} \bibinfo{person}{Klaus
  Mueller}.} \bibinfo{year}{2021}\natexlab{}.
\newblock \showarticletitle{Fluent: An AI Augmented Writing Tool for People who
  Stutter}. In \bibinfo{booktitle}{\emph{The 23rd International ACM SIGACCESS
  Conference on Computers and Accessibility}}. \bibinfo{publisher}{ACM},
  \bibinfo{pages}{1--8}.
\newblock


\bibitem[Goberman et~al\mbox{.}(2010)]%
        {goberman2010characteristics}
\bibfield{author}{\bibinfo{person}{Alexander~M Goberman},
  \bibinfo{person}{Michael Blomgren}, {and} \bibinfo{person}{Erika Metzger}.}
  \bibinfo{year}{2010}\natexlab{}.
\newblock \showarticletitle{Characteristics of speech disfluency in Parkinson
  disease}.
\newblock \bibinfo{journal}{\emph{Journal of Neurolinguistics}}
  \bibinfo{volume}{23}, \bibinfo{number}{5} (\bibinfo{year}{2010}),
  \bibinfo{pages}{470--478}.
\newblock


\bibitem[Gondala et~al\mbox{.}(2021)]%
        {gondala2021error}
\bibfield{author}{\bibinfo{person}{Sashank Gondala}, \bibinfo{person}{Lyan
  Verwimp}, \bibinfo{person}{Ernest Pusateri}, \bibinfo{person}{Manos
  Tsagkias}, {and} \bibinfo{person}{Christophe Van~Gysel}.}
  \bibinfo{year}{2021}\natexlab{}.
\newblock \showarticletitle{Error-driven Pruning of Language Models for Virtual
  Assistants}.
\newblock \bibinfo{journal}{\emph{arXiv preprint arXiv:2102.07219}}
  (\bibinfo{year}{2021}).
\newblock


\bibitem[Graves(2012)]%
        {graves2012RNNT}
\bibfield{author}{\bibinfo{person}{Alex Graves}.}
  \bibinfo{year}{2012}\natexlab{}.
\newblock \showarticletitle{Sequence transduction with recurrent neural
  networks}.
\newblock \bibinfo{journal}{\emph{arXiv preprint arXiv:1211.3711}}
  (\bibinfo{year}{2012}).
\newblock


\bibitem[Green et~al\mbox{.}(2021)]%
        {green2021automatic}
\bibfield{author}{\bibinfo{person}{Jordan~R Green}, \bibinfo{person}{Robert~L
  MacDonald}, \bibinfo{person}{Pan-Pan Jiang}, \bibinfo{person}{Julie Cattiau},
  \bibinfo{person}{Rus Heywood}, \bibinfo{person}{Richard Cave},
  \bibinfo{person}{Katie Seaver}, \bibinfo{person}{Marilyn~A Ladewig},
  \bibinfo{person}{Jimmy Tobin}, \bibinfo{person}{Michael~P Brenner},
  {et~al\mbox{.}}} \bibinfo{year}{2021}\natexlab{}.
\newblock \showarticletitle{Automatic Speech Recognition of Disordered Speech:
  Personalized Models Outperforming Human Listeners on Short Phrases.}. In
  \bibinfo{booktitle}{\emph{Interspeech}}. \bibinfo{pages}{4778--4782}.
\newblock


\bibitem[Heafield et~al\mbox{.}(2013)]%
        {Heafield2013LM}
\bibfield{author}{\bibinfo{person}{Kenneth Heafield}, \bibinfo{person}{Ivan
  Pouzyrevsky}, \bibinfo{person}{Jonathan Clark}, {and}
  \bibinfo{person}{Philipp Koehn}.} \bibinfo{year}{2013}\natexlab{}.
\newblock \showarticletitle{Scalable Modified Kneser-Ney Language Model
  Estimation}. In \bibinfo{booktitle}{\emph{In Proceedings of the 51st Annual
  Meeting of the Association for Computational Linguistics (Volume 2: Short
  Papers), pages 690–696}}. Association for Computational Linguistics.
\newblock


\bibitem[Heeman et~al\mbox{.}(2016)]%
        {HeemanInterSpeech16}
\bibfield{author}{\bibinfo{person}{Peter~A Heeman}, \bibinfo{person}{Rebecca
  Lunsford}, \bibinfo{person}{Andy McMillin}, {and} \bibinfo{person}{J~Scott
  Yaruss}.} \bibinfo{year}{2016}\natexlab{}.
\newblock \showarticletitle{Using Clinician Annotations to Improve Automatic
  Speech Recognition of Stuttered Speech}. In
  \bibinfo{booktitle}{\emph{Interspeech}}. \bibinfo{pages}{2651--2655}.
\newblock


\bibitem[Hinton et~al\mbox{.}(2012)]%
        {hinton2012HybridASR}
\bibfield{author}{\bibinfo{person}{Geoffrey Hinton}, \bibinfo{person}{Li Deng},
  \bibinfo{person}{Dong Yu}, \bibinfo{person}{George~E Dahl},
  \bibinfo{person}{Abdel-rahman Mohamed}, \bibinfo{person}{Navdeep Jaitly},
  \bibinfo{person}{Andrew Senior}, \bibinfo{person}{Vincent Vanhoucke},
  \bibinfo{person}{Patrick Nguyen}, \bibinfo{person}{Tara~N Sainath},
  {et~al\mbox{.}}} \bibinfo{year}{2012}\natexlab{}.
\newblock \showarticletitle{Deep neural networks for acoustic modeling in
  speech recognition: The shared views of four research groups}.
\newblock \bibinfo{journal}{\emph{IEEE Signal processing magazine}}
  \bibinfo{volume}{29}, \bibinfo{number}{6} (\bibinfo{year}{2012}),
  \bibinfo{pages}{82--97}.
\newblock


\bibitem[Howell et~al\mbox{.}(2009)]%
        {UCLASS}
\bibfield{author}{\bibinfo{person}{Peter Howell}, \bibinfo{person}{Stephen
  Davis}, {and} \bibinfo{person}{Jon Bartrip}.}
  \bibinfo{year}{2009}\natexlab{}.
\newblock \showarticletitle{The {UCLASS} archive of stuttered speech}.
\newblock \bibinfo{journal}{\emph{Journal of Speech Language and Hearing
  Research}}  \bibinfo{volume}{52} (\bibinfo{year}{2009}),
  \bibinfo{pages}{556}.
\newblock


\bibitem[Huang et~al\mbox{.}(2020)]%
        {huang2020sndcnn}
\bibfield{author}{\bibinfo{person}{Zhen Huang}, \bibinfo{person}{Tim Ng},
  \bibinfo{person}{Leo Liu}, \bibinfo{person}{Henry Mason},
  \bibinfo{person}{Xiaodan Zhuang}, {and} \bibinfo{person}{Daben Liu}.}
  \bibinfo{year}{2020}\natexlab{}.
\newblock \showarticletitle{SNDCNN: Self-normalizing deep CNNs with scaled
  exponential linear units for speech recognition}. In
  \bibinfo{booktitle}{\emph{ICASSP 2020-2020 IEEE International Conference on
  Acoustics, Speech and Signal Processing (ICASSP)}}. IEEE,
  \bibinfo{pages}{6854--6858}.
\newblock


\bibitem[Juste et~al\mbox{.}(2018)]%
        {juste2018frequency}
\bibfield{author}{\bibinfo{person}{Fabiola~Star{\'o}bole Juste},
  \bibinfo{person}{Fernanda~Chiarion Sassi}, \bibinfo{person}{Julia~Biancalana
  Costa}, {and} \bibinfo{person}{Claudia Regina~Furquim de Andrade}.}
  \bibinfo{year}{2018}\natexlab{}.
\newblock \showarticletitle{Frequency of speech disruptions in Parkinson's
  Disease and developmental stuttering: A comparison among speech tasks}.
\newblock \bibinfo{journal}{\emph{Plos one}} \bibinfo{volume}{13},
  \bibinfo{number}{6} (\bibinfo{year}{2018}), \bibinfo{pages}{e0199054}.
\newblock


\bibitem[Kourkounakis et~al\mbox{.}(2020)]%
        {kourkounakis2020detecting}
\bibfield{author}{\bibinfo{person}{Tedd Kourkounakis},
  \bibinfo{person}{Amirhossein Hajavi}, {and} \bibinfo{person}{Ali Etemad}.}
  \bibinfo{year}{2020}\natexlab{}.
\newblock \showarticletitle{Detecting Multiple Speech Disfluencies Using a Deep
  Residual Network with Bidirectional Long Short-Term Memory}. In
  \bibinfo{booktitle}{\emph{ICASSP}}. IEEE.
\newblock


\bibitem[Le et~al\mbox{.}(2020)]%
        {le2020hybridASR}
\bibfield{author}{\bibinfo{person}{Duc Le}, \bibinfo{person}{Thilo Koehler},
  \bibinfo{person}{Christian Fuegen}, {and} \bibinfo{person}{Michael~L.
  Seltzer}.} \bibinfo{year}{2020}\natexlab{}.
\newblock \showarticletitle{G2G: TTS-DRIVEN PRONUNCIATION LEARNING FOR
  GRAPHEMIC HYBRID ASR}. In \bibinfo{booktitle}{\emph{ICASSP 2020-2020 IEEE
  International Conference on Acoustics, Speech and Signal Processing
  (ICASSP)}}. IEEE, \bibinfo{pages}{6869--6873}.
\newblock


\bibitem[Lea et~al\mbox{.}(2021a)]%
        {lea2021sep}
\bibfield{author}{\bibinfo{person}{Colin Lea}, \bibinfo{person}{Vikramjit
  Mitra}, \bibinfo{person}{Aparna Joshi}, \bibinfo{person}{Sachin Kajarekar},
  {and} \bibinfo{person}{Jeffrey~P Bigham}.} \bibinfo{year}{2021}\natexlab{a}.
\newblock \showarticletitle{SEP-28k: A Dataset for Stuttering Event Detection
  From Podcasts With People Who Stutter}.
\newblock \bibinfo{journal}{\emph{arXiv preprint arXiv:2102.12394}}
  (\bibinfo{year}{2021}).
\newblock


\bibitem[Lea et~al\mbox{.}(2021b)]%
        {LeaStutterDetection2021}
\bibfield{author}{\bibinfo{person}{Colin Lea}, \bibinfo{person}{Vikramjit
  Mitra}, \bibinfo{person}{Aparna Joshi}, \bibinfo{person}{Sachin Kajarekar},
  {and} \bibinfo{person}{Jeffrey~P. Bigham}.} \bibinfo{year}{2021}\natexlab{b}.
\newblock \showarticletitle{SEP-28k: A Dataset for Stuttering Event Detection
  From Podcasts With People Who Stutter}. In
  \bibinfo{booktitle}{\emph{ICASSP}}. IEEE.
\newblock


\bibitem[Li et~al\mbox{.}(2020)]%
        {li2020RNNTEndpointer}
\bibfield{author}{\bibinfo{person}{Bo Li}, \bibinfo{person}{Shuo-yiin Chang},
  \bibinfo{person}{Tara~N Sainath}, \bibinfo{person}{Ruoming Pang},
  \bibinfo{person}{Yanzhang He}, \bibinfo{person}{Trevor Strohman}, {and}
  \bibinfo{person}{Yonghui Wu}.} \bibinfo{year}{2020}\natexlab{}.
\newblock \showarticletitle{Towards fast and accurate streaming end-to-end
  ASR}. In \bibinfo{booktitle}{\emph{ICASSP 2020-2020 IEEE International
  Conference on Acoustics, Speech and Signal Processing (ICASSP)}}. IEEE,
  \bibinfo{pages}{6069--6073}.
\newblock


\bibitem[Maas et~al\mbox{.}(2018)]%
        {Maas2018DNNEndpointer}
\bibfield{author}{\bibinfo{person}{Roland Maas}, \bibinfo{person}{Ariya
  Rastrow}, \bibinfo{person}{Chengyuan Ma}, \bibinfo{person}{Guitang Lan},
  \bibinfo{person}{Kyle Goehner}, \bibinfo{person}{Gautam Tiwari},
  \bibinfo{person}{Shaun Joseph}, {and} \bibinfo{person}{Björn Hoffmeister}.}
  \bibinfo{year}{2018}\natexlab{}.
\newblock \showarticletitle{Combining Acoustic Embeddings and Decoding Features
  for End-of-Utterance Detection in Real-Time Far-Field Speech Recognition
  Systems}. In \bibinfo{booktitle}{\emph{2018 IEEE International Conference on
  Acoustics, Speech and Signal Processing (ICASSP)}}.
  \bibinfo{pages}{5544--5548}.
\newblock


\bibitem[MacDonald et~al\mbox{.}(2021)]%
        {projecteuphonia}
\bibfield{author}{\bibinfo{person}{Bob MacDonald}, \bibinfo{person}{Pan-Pan
  Jiang}, \bibinfo{person}{Julie Cattiau}, \bibinfo{person}{Rus Heywood},
  \bibinfo{person}{Richard Cave}, \bibinfo{person}{Katie Seaver},
  \bibinfo{person}{Marilyn Ladewig}, \bibinfo{person}{Jimmy Tobin},
  \bibinfo{person}{Michael Brenner}, \bibinfo{person}{Philip~Q Nelson},
  \bibinfo{person}{Jordan~R. Green}, {and} \bibinfo{person}{Katrin Tomanek}.}
  \bibinfo{year}{2021}\natexlab{}.
\newblock \showarticletitle{Disordered Speech Data Collection: Lessons Learned
  at 1 Million Utterances from Project Euphonia}.
\newblock


\bibitem[Madeira et~al\mbox{.}(2013)]%
        {madeira2013building}
\bibfield{author}{\bibinfo{person}{Rui~Neves Madeira},
  \bibinfo{person}{Patr{\'\i}cia Macedo}, \bibinfo{person}{Pedro Pita},
  \bibinfo{person}{{\'I}ris Bonan{\c{c}}a}, {and} \bibinfo{person}{Helena
  Germano}.} \bibinfo{year}{2013}\natexlab{}.
\newblock \showarticletitle{Building on mobile towards better stuttering
  awareness to improve speech therapy}. In
  \bibinfo{booktitle}{\emph{Proceedings of International Conference on Advances
  in Mobile Computing \& Multimedia}}. \bibinfo{pages}{551--554}.
\newblock


\bibitem[Mahesha and Vinod(2016)]%
        {Mahesha2016}
\bibfield{author}{\bibinfo{person}{P. Mahesha} {and} \bibinfo{person}{D.S.
  Vinod}.} \bibinfo{year}{2016}\natexlab{}.
\newblock \showarticletitle{Gaussian Mixture Model Based Classification of
  Stuttering Dysfluencies}.
\newblock \bibinfo{journal}{\emph{Journal of Intelligent Systems}}
  \bibinfo{volume}{25}, \bibinfo{number}{3} (\bibinfo{year}{2016}),
  \bibinfo{pages}{387--399}.
\newblock


\bibitem[McNaney et~al\mbox{.}(2018)]%
        {mcnaney2018stammerapp}
\bibfield{author}{\bibinfo{person}{Roisin McNaney},
  \bibinfo{person}{Christopher Bull}, \bibinfo{person}{Lynne Mackie},
  \bibinfo{person}{Floriane Dahman}, \bibinfo{person}{Helen Stringer},
  \bibinfo{person}{Dan Richardson}, {and} \bibinfo{person}{Daniel Welsh}.}
  \bibinfo{year}{2018}\natexlab{}.
\newblock \showarticletitle{StammerApp: Designing a Mobile Application to
  Support Self-Reflection and Goal Setting for People Who Stammer}. In
  \bibinfo{booktitle}{\emph{Proceedings of the 2018 CHI Conference on Human
  Factors in Computing Systems}}. \bibinfo{publisher}{ACM},
  \bibinfo{pages}{1--12}.
\newblock


\bibitem[Mendelev et~al\mbox{.}(2020)]%
        {Mendelev2020}
\bibfield{author}{\bibinfo{person}{Valentin Mendelev}, \bibinfo{person}{Tina
  Raissi}, \bibinfo{person}{Guglielmo Camporese}, {and} \bibinfo{person}{Manuel
  Giollo}.} \bibinfo{year}{2020}\natexlab{}.
\newblock \showarticletitle{Improved Robustness to Disfluencies in
  RNN-Transducer Based Speech Recognition}. In
  \bibinfo{booktitle}{\emph{arXiv}}.
\newblock
\showeprint{2012.06259}~[cs.CL]


\bibitem[Mitra et~al\mbox{.}(2021)]%
        {mitra2021}
\bibfield{author}{\bibinfo{person}{Vikramjit Mitra}, \bibinfo{person}{Zifang
  Huang}, \bibinfo{person}{Colin Lea}, \bibinfo{person}{Lauren Tooley},
  \bibinfo{person}{Sarah Wu}, \bibinfo{person}{Darren Botten},
  \bibinfo{person}{Ashwini Palekar}, \bibinfo{person}{Shrinath Thelapurath},
  \bibinfo{person}{Panayiotis Georgiou}, \bibinfo{person}{Sachin Kajarekar},
  {et~al\mbox{.}}} \bibinfo{year}{2021}\natexlab{}.
\newblock \showarticletitle{Analysis and tuning of a voice assistant system for
  dysfluent speech}.
\newblock \bibinfo{journal}{\emph{arXiv preprint arXiv:2106.11759}}
  (\bibinfo{year}{2021}).
\newblock


\bibitem[Murry(2014)]%
        {murry2014spasmodic}
\bibfield{author}{\bibinfo{person}{Thomas Murry}.}
  \bibinfo{year}{2014}\natexlab{}.
\newblock \showarticletitle{Spasmodic dysphonia: let's look at that again}.
\newblock \bibinfo{journal}{\emph{Journal of Voice}} \bibinfo{volume}{28},
  \bibinfo{number}{6} (\bibinfo{year}{2014}), \bibinfo{pages}{694--699}.
\newblock


\bibitem[{National Public Media}(2022)]%
        {npr_smart_audio}
\bibfield{author}{\bibinfo{person}{{National Public Media}}.}
  \bibinfo{year}{2022}\natexlab{}.
\newblock \bibinfo{title}{NPR Smart Audio Report}.
\newblock
  \bibinfo{howpublished}{\url{https://www.nationalpublicmedia.com/insights/reports/smart-audio-report/}}.
\newblock


\bibitem[Pradhan et~al\mbox{.}(2018)]%
        {pradhan2018accessibility}
\bibfield{author}{\bibinfo{person}{Alisha Pradhan}, \bibinfo{person}{Kanika
  Mehta}, {and} \bibinfo{person}{Leah Findlater}.}
  \bibinfo{year}{2018}\natexlab{}.
\newblock \showarticletitle{``Accessibility Came by Accident'': Use of
  Voice-Controlled Intelligent Personal Assistants by People with
  Disabilities}. In \bibinfo{booktitle}{\emph{Proceedings of the 2018 CHI
  Conference on human factors in computing systems}}. \bibinfo{pages}{1--13}.
\newblock


\bibitem[Ratnera and MacWhinney(2018)]%
        {FluencyBank}
\bibfield{author}{\bibinfo{person}{Nan~Bernstein Ratnera} {and}
  \bibinfo{person}{Brian MacWhinney}.} \bibinfo{year}{2018}\natexlab{}.
\newblock \showarticletitle{Fluency Bank: A new resource for fluency research
  and practice}.
\newblock \bibinfo{journal}{\emph{Journal of Fluency Disorders}}
  (\bibinfo{year}{2018}).
\newblock


\bibitem[Riad et~al\mbox{.}(2020)]%
        {riad2020identification}
\bibfield{author}{\bibinfo{person}{Rachid Riad},
  \bibinfo{person}{Anne-Catherine Bachoud-L{\'e}vi}, \bibinfo{person}{Frank
  Rudzicz}, {and} \bibinfo{person}{Emmanuel Dupoux}.}
  \bibinfo{year}{2020}\natexlab{}.
\newblock \showarticletitle{Identification of Primary and Collateral Tracks in
  Stuttered Speech}. In \bibinfo{booktitle}{\emph{LREC}}.
  \bibinfo{publisher}{European Language Resources Association}.
\newblock


\bibitem[Riley(2009)]%
        {SSI}
\bibfield{author}{\bibinfo{person}{Glyndon~D Riley}.}
  \bibinfo{year}{2009}\natexlab{}.
\newblock \bibinfo{booktitle}{\emph{{SSI-4} stuttering severity instrument
  fourth edition}}.
\newblock


\bibitem[Roper et~al\mbox{.}(2019)]%
        {roper2019speech}
\bibfield{author}{\bibinfo{person}{Abi Roper}, \bibinfo{person}{Stephanie
  Wilson}, \bibinfo{person}{Timothy Neate}, {and} \bibinfo{person}{Jane
  Marshall}.} \bibinfo{year}{2019}\natexlab{}.
\newblock \showarticletitle{Speech and Language}.
\newblock In \bibinfo{booktitle}{\emph{Web Accessibility}}.
  \bibinfo{publisher}{Springer}, \bibinfo{pages}{121--131}.
\newblock


\bibitem[Sak et~al\mbox{.}(2013)]%
        {Sak2013HybridModelComponents}
\bibfield{author}{\bibinfo{person}{Hasim Sak}, \bibinfo{person}{Francoise
  Beaufays}, \bibinfo{person}{Kaisuke Nakajima}, {and} \bibinfo{person}{Cyril
  Allauzen}.} \bibinfo{year}{2013}\natexlab{}.
\newblock \showarticletitle{Language model verbalization for automatic speech
  recognition}. In \bibinfo{booktitle}{\emph{ICASSP 2013 IEEE International
  Conference on Acoustics, Speech and Signal Processing (ICASSP)}}. IEEE.
\newblock


\bibitem[Sander(1963)]%
        {sander1963frequency}
\bibfield{author}{\bibinfo{person}{Eric~K Sander}.}
  \bibinfo{year}{1963}\natexlab{}.
\newblock \showarticletitle{Frequency of syllable repetition and
  ‘stutterer’judgments}.
\newblock \bibinfo{journal}{\emph{Journal of Speech and Hearing Disorders}}
  \bibinfo{volume}{28}, \bibinfo{number}{1} (\bibinfo{year}{1963}),
  \bibinfo{pages}{19--30}.
\newblock


\bibitem[Shonibare et~al\mbox{.}(2022)]%
        {shonibare2022enhancing}
\bibfield{author}{\bibinfo{person}{Olabanji Shonibare}, \bibinfo{person}{Xiaosu
  Tong}, {and} \bibinfo{person}{Venkatesh Ravichandran}.}
  \bibinfo{year}{2022}\natexlab{}.
\newblock \showarticletitle{Enhancing ASR for Stuttered Speech with Limited
  Data Using Detect and Pass}.
\newblock \bibinfo{journal}{\emph{arXiv preprint arXiv:2202.05396}}
  (\bibinfo{year}{2022}).
\newblock


\bibitem[Stolcke and Droppo(2017)]%
        {Stolcke2017HumanMachineComparison}
\bibfield{author}{\bibinfo{person}{Andreas Stolcke} {and}
  \bibinfo{person}{Jasha Droppo}.} \bibinfo{year}{2017}\natexlab{}.
\newblock \showarticletitle{Comparing Human and Machine Errors in
  Conversational Speech Transcription}. In \bibinfo{booktitle}{\emph{Proc of
  INTERSPEECH 2017}}. IEEE, \bibinfo{pages}{137--141}.
\newblock


\bibitem[Subtelny et~al\mbox{.}(1992)]%
        {subtelny1992spectral}
\bibfield{author}{\bibinfo{person}{Joanne~D Subtelny},
  \bibinfo{person}{Robert~L Whitehead}, {and} \bibinfo{person}{Vincent~J
  Samar}.} \bibinfo{year}{1992}\natexlab{}.
\newblock \showarticletitle{Spectral study of deviant resonance in the speech
  of women who are deaf}.
\newblock \bibinfo{journal}{\emph{Journal of Speech, Language, and Hearing
  Research}} \bibinfo{volume}{35}, \bibinfo{number}{3} (\bibinfo{year}{1992}),
  \bibinfo{pages}{574--579}.
\newblock


\bibitem[Tichenor and Yaruss(2021)]%
        {tichenor2021variability}
\bibfield{author}{\bibinfo{person}{Seth~E Tichenor} {and}
  \bibinfo{person}{J~Scott Yaruss}.} \bibinfo{year}{2021}\natexlab{}.
\newblock \showarticletitle{Variability of stuttering: Behavior and impact}.
\newblock \bibinfo{journal}{\emph{American Journal of Speech-Language
  Pathology}} \bibinfo{volume}{30}, \bibinfo{number}{1} (\bibinfo{year}{2021}),
  \bibinfo{pages}{75--88}.
\newblock


\bibitem[Tobin and Tomanek(2022)]%
        {tobin2022personalized}
\bibfield{author}{\bibinfo{person}{Jimmy Tobin} {and} \bibinfo{person}{Katrin
  Tomanek}.} \bibinfo{year}{2022}\natexlab{}.
\newblock \showarticletitle{Personalized Automatic Speech Recognition Trained
  on Small Disordered Speech Datasets}. In \bibinfo{booktitle}{\emph{ICASSP
  2022-2022 IEEE International Conference on Acoustics, Speech and Signal
  Processing (ICASSP)}}. IEEE, \bibinfo{pages}{6637--6641}.
\newblock


\bibitem[Ward and Scott(2011)]%
        {ward2011cluttering}
\bibfield{author}{\bibinfo{person}{David Ward} {and}
  \bibinfo{person}{Kathleen~S Scott}.} \bibinfo{year}{2011}\natexlab{}.
\newblock \showarticletitle{Cluttering: A Handbook of Research}.
\newblock \bibinfo{journal}{\emph{Intervention and Education}}
  (\bibinfo{year}{2011}).
\newblock


\bibitem[Wheeler(2020)]%
        {USAToday}
\bibfield{author}{\bibinfo{person}{Kevin Wheeler}.}
  \bibinfo{year}{2020}\natexlab{}.
\newblock \bibinfo{title}{For people who stutter, the convenience of voice
  assistant technology remains out of reach}.
\newblock \bibinfo{howpublished}{USA Today (online)}.
\newblock


\bibitem[Xiong et~al\mbox{.}(2016)]%
        {Xiong2016HumanParityASR}
\bibfield{author}{\bibinfo{person}{Wayne Xiong}, \bibinfo{person}{Jasha
  Droppo}, \bibinfo{person}{Xuedong Huang}, \bibinfo{person}{Frank Seide},
  \bibinfo{person}{Mike Seltzer}, \bibinfo{person}{Andreas Stolcke},
  \bibinfo{person}{Dong Yu}, {and} \bibinfo{person}{Geoffrey Zweig}.}
  \bibinfo{year}{2016}\natexlab{}.
\newblock \showarticletitle{Achieving Human Parity in Conversational Speech
  Recognition}.
\newblock \bibinfo{journal}{\emph{CoRR}}  \bibinfo{volume}{abs/1610.05256}.
\newblock
\showeprint[arXiv]{1610.05256}


\bibitem[Xiong et~al\mbox{.}(2018)]%
        {Xiong2018MicrosoftASR}
\bibfield{author}{\bibinfo{person}{W. Xiong}, \bibinfo{person}{L. Wu},
  \bibinfo{person}{F. Alleva}, \bibinfo{person}{J. Droppo}, \bibinfo{person}{X.
  Huang}, {and} \bibinfo{person}{A. Stolcke}.} \bibinfo{year}{2018}\natexlab{}.
\newblock \showarticletitle{The Microsoft 2017 Conversational Speech
  Recognition System}. In \bibinfo{booktitle}{\emph{{IEEE} International
  Conference on Acoustics, Speech and Signal Processing ({ICASSP})}}.
  \bibinfo{publisher}{{IEEE}}.
\newblock


\bibitem[Yaruss and Quesal(2006)]%
        {yaruss2006overall}
\bibfield{author}{\bibinfo{person}{J~Scott Yaruss} {and}
  \bibinfo{person}{Robert~W Quesal}.} \bibinfo{year}{2006}\natexlab{}.
\newblock \showarticletitle{Overall Assessment of the Speaker's Experience of
  Stuttering (OASES): Documenting multiple outcomes in stuttering treatment}.
\newblock \bibinfo{journal}{\emph{Journal of fluency disorders}}
  \bibinfo{volume}{31}, \bibinfo{number}{2} (\bibinfo{year}{2006}),
  \bibinfo{pages}{90--115}.
\newblock


\bibitem[Yi et~al\mbox{.}(2021)]%
        {yi2021hybridASR}
\bibfield{author}{\bibinfo{person}{Cheng Yi}, \bibinfo{person}{Shiyu Zhou},
  {and} \bibinfo{person}{Bo Xu}.} \bibinfo{year}{2021}\natexlab{}.
\newblock \showarticletitle{Effciently fusing pretrained acoustic and
  linguistic encoders for low-resource speech recognition}. In
  \bibinfo{booktitle}{\emph{IEEE Signal Processing Letters}},
  Vol.~\bibinfo{volume}{28}. IEEE, \bibinfo{pages}{788--792}.
\newblock


\bibitem[Zhong et~al\mbox{.}(2014)]%
        {zhong2014justspeak}
\bibfield{author}{\bibinfo{person}{Yu Zhong}, \bibinfo{person}{TV Raman},
  \bibinfo{person}{Casey Burkhardt}, \bibinfo{person}{Fadi Biadsy}, {and}
  \bibinfo{person}{Jeffrey~P Bigham}.} \bibinfo{year}{2014}\natexlab{}.
\newblock \showarticletitle{JustSpeak: enabling universal voice control on
  Android}. In \bibinfo{booktitle}{\emph{Proceedings of the 11th Web for All
  Conference}}. \bibinfo{pages}{1--4}.
\newblock


\end{thebibliography}

\appendix


\section*{Supplementary material}

# Survey for People Who Stutter

**Study Introduction** We are studying experiences with using speech to control digital devices such as phones, watches, smart speakers, tablets, computers, and car infotainment systems. For example, a person could ask their smart speaker or phone "What is the time?" or "Set a timer for 5 minutes", could tell their phone or watch to "Call Mom", or could dictate rather than type longer-form content like a text message, email or document. We are interested in responses regardless of how much you currently use these systems and regardless of how well these systems work for you right now. Please do not submit any personally identifiable, sensitive, or confidential information in your responses.

## Participant Background

Q: What is your gender?

- Man
- Woman
- Non-binary
- Prefer not to disclose
- Prefer to self-describe:(fill in)

Q: What is your age range?

- 18-19
- 20-29
- 30-39
- 40-49
- 50-59
- 60-69
- 70-79
- 80+

Q: What is your ethnicity?

- Hispanic, Latino, or of Spanish origin
- American Indian or Alaska Native
- Asian
- Black or African American
- Native Hawaiian or Other Pacific Islander
- White
- Prefer not to answer

Q: How often do immediate family members, close friends, or caregivers understand your speech?

- Never
- Rarely
- Sometimes

- Usually
- Always

Q: How often do acquaintances or people you don't talk to often understand your speech?

- Never
- Rarely
- Sometimes
- Usually
- Always

Q: How often do strangers understand your speech?

- Never
- Rarely
- Sometimes
- Usually
- Always

Q: How much do your speech characteristics vary over time? Select all that apply.

- Always the same
- Varies throughout the day
- Varies day-to-day
- Changes progressively over weeks or months

Q: Would you describe your speech with any of the following?

- Breathy
- Strained
- Monotone
- Intermittent prolongations
- Slurred (distorted vowels)
- Frequent repetitions
- Frequent filler words (e.g., "uh", "um", "you know")
- Frequent blocks/pauses
- Low/Quiet volume
- Nasal
- Other (please specify)

Q: How much physical effort does it take for you to speak?

- Very high effort
- Some effort
- Neutral
- Low effort
- Very low

Q: Which of the following devices do you own or are in your living space? (Only select devices where you could imagine using a voice assistant)

- Smartphone (fill in type)
- Smart Speaker (fill in type)
- Smart Watch (fill in type)
- Tablet (fill in type)
- Computer (fill in type)

Q: How do you access your phone, tablet, or computer? (Select all that apply)

- Direct access with hands - touching the screen
- Voice/speech control
- Standard keyboard
- Standard mouse
- Head tracking
- Head stick/stylus
- Mouth stick/stylus
- Eye tracking
- Scanning/switch control
- Someone helps me
- Other - please describe

## Voice Assistants

Voice assistants listen for spoken questions or commands. They usually have a spoken response but some can also show visual information. Examples are Amazon Alexa, Apple Siri, and Google Assistant. They can be found on smart speakers, like Amazon Echo or Apple HomePod devices, or on phones and computers.

Q: Have you heard of voice assistants or observed other people using them?

- Yes
- No

If you have never heard of a voice assistant then skip to Q38.

Q: Have you ever tried using a voice assistant yourself ?

- Yes
- No

If you have never tried using a voice assistant then skip to Q38.

Q: How often do you currently use a voice assistant?

- At least daily
- At least weekly
- At least monthly
- Less than monthly

- Never
- I don't know

Q: How useful are voice assistants for you currently? and why?

- Very useful
- Somewhat useful
- Not very useful
- Not at all useful

Q: How often do voice assistants recognize your speech accurately?

- Always
- Often
- Sometimes
- Rarely
- Never

Q: On what devices do you or have you previously used a voice assistant?* Check all that apply.*

- Phone
- Watch
- Tablet
- Computer
- Smart speaker
- Car infotainment system
- Other (please specify)

Q: For what tasks have you used a voice assistant? *(select all that apply)*

- Weather
- Music
- Asking general questions
- Setting Alarms
- Setting Timers
- Checking Time
- Making Phone calls
- Using Shortcuts
- HomeKit
- Web search and browsing
- Create reminders/todo list
- Sending messages
- Get the news
- Read or listen to messages
- Dictionary
- Not interested in any
- Other (please specify)

Q: If voice assistants worked better for you, what tasks would you like to use a voice assistant? *(select all that apply)*

- Weather
- Music
- Asking general questions
- Setting Alarms
- Setting Timers
- Checking Time
- Making Phone calls
- Using Shortcuts
- HomeKit
- Web search and browsing
- Create reminders/todo list
- Sending messages
- Get the news
- Read or listen to messages
- Dictionary
- Not interested in any
- Other (please specify)

Q: Do any of the following reasons prevent you from using a voice assistant more often? If yes, select all that apply:

- Voice Assistants work most or all of the time
- I don't have any devices that include a voice assistant
- Voice assistants don't seem useful
- It takes too much physical effort for me to speak
- Voice assistants don't understand my speech well enough
- I don't want other people to hear me talking to the voice assistant
- Voice Assistants cut off my speech before I finish talking
- Other (please specify)

Q: When using a voice assistant, how often do you revise or re-state what you say due to inaccuracies in speech recognition?

- Always
- Often
- Sometimes
- Rarely
- Never
- I don't use a voice assistant

Q: If you revise/re-state what you say, is the command typically recognized on subsequent attempt(s)?

- Always
- Often
- Sometimes

- Rarely
- Never
- I don't typically do this

Q: Most voice assistants use a "wake" phrase such as "Ok Google," "Alexa," or "Hey Siri" to start the system. How often do these phrases work for you?

- Always
- Often
- Sometimes
- Rarely
- Never
- I haven't used a voice assistant with a "wake" word

Q: With some devices, you can press down a button while you're speaking to the voice assistant and only release thebutton when you're done talking. This means the voice assistant doesn't have to automatically guess when you're done speaking. Have you used a device that lets you press and hold a button while you're speaking to the voice assistant?

- Yes
- No
- I don't know

Q: *If yes:* How useful is it to be able to press down and hold the button while you're speaking, so that the voice assistant doesn't need to automatically guess when you're done speaking? Why?

- Very useful
- Useful
- Not very useful
- Not at all useful

Q: Voice assistants automatically guess when the person is done speaking. How often do voice assistants or dictation tools cut you off before you're done speaking?

- Always
- Often
- Sometimes
- Rarely
- Never

Q: Do you know why they cut you off? (fill in)

Q: Have you tried using a keyboard to type commands to voice assistants?

- Yes
- No
- I don't know what this is

Q: If yes to above, how useful is it for you to type instead of speaking to a voice assistant? Why?

- Very useful
- Useful
- Not very useful
- Not at all useful

Q: If you haven't used a voice assistant before, why haven't you tried it? Select all that apply:

- I don't know what voice assistants are
- I don't have any devices that include a voice assistant
- Voice assistants don't seem useful
- It takes too much physical effort for me to speak
- Voice assistants don't understand my speech well enough
- I don't want other people to hear me talking to the voice assistant
- Other (please specify)

Q: How often do you expect voice assistants would be able to understand your speech correctly?

- Always
- Often
- Sometimes
- Rarely
- Never

Q: If you haven't used a voice assistant before, is there anything else you want to add about why you haven't tried?

Q: If you don't regularly use a voice assistant then what, if anything, would make you start?

## Dictation

With voice dictation, a phone, computer or other device automatically recognizes a person's speech and inserts that text into a text message, note, document, or other app. For example, you could respond to a text while with your voice while in a car. Dictation provides an alternative to using a keyboard to enter text. This does not include tasks like "set a timer."

Q: Have you heard of voice dictation or observed other people using it?

- Yes
- No

**If you have never heard of Dictation then skip to Q54.**

Q. Have you ever tried using dictation yourself?

- Yes

- No

**If you have never tried using Dictation then skip to Q54.**

Q: How often do you currently use dictation?

- At least daily
- At least weekly
- At least monthly
- Less than monthly
- Never
- I don't know

Q: On what devices do you or have you previously used dictation? *Check all that apply.*

- Phone
- Watch
- Tablet
- Computer
- Smart speaker
- Car infotainment system
- Other (please specify)

Q: For what tasks have you used dictation? *(select all that apply)*

- To write a message via text or other instant messenger
- To take notes
- To write an email
- In a web browser
- None
- Other (please specify)

Q: If dictation worked better for you, what tasks would you like to use it for? *(select all that apply)*

- To write a message via text or other instance messenger (e.g., Facebook Messenger)
- To take notes, to write an email
- In a web browser
- None
- Other (please specify)

Q: Do any of the following reasons prevent you from using dictation more often? If yes, select all that apply:

- I don't have any devices that support dictation
- Dictation doesn't seem useful
- It takes too much physical effort for me to speak
- Dictation tools don't understand my speech well enough
- I don't want other people to hear me dictate to the device

- I use more refined language when I type than when I speak
- Other (please specify)

Q: When using dictation, how often do you revise or re-state what you say due to inaccuracies in speech recognition?

- Always
- Often
- Sometimes
- Rarely
- Never
- I don't use dictation

Q: If you revise/re-state what you say, is the text typically recognized on subsequent attempt(s)?

- Always
- Often
- Sometimes
- Rarely
- I don't typically do this

Q: How often does dictation recognize your speech accurately?

- Always
- Often
- Sometimes
- Rarely
- Never

Q: How useful is dictation for you currently? Why?

- Very useful
- Somewhat useful
- Not very useful
- Not at all useful

Q: If you don't regularly use dictation then what, if anything, would make you start?

**If no haven't used speech at all or haven't used dictation specifically...**

Q: Why haven't you tried using dictation? Select all that apply:

- I don't know what dictation is
- I don't have any devices that support dictation
- Dictation doesn't seem useful
- It takes too much effort for me to speak
- Dictation tools don't understand my speech well enough
- I don't want other people to hear me dictate to the device
- I use more refined language when I type than when I speak
- Other (please specify)

Q: How often do you expect voice dictation would be able to understand your speech correctly? *..Why?*

- Always
- Often
- Sometimes
- Rarely
- Never

Q: What, if anything, would make you interested in using dictation?

Q: Is there anything else you want to add about why you haven't tried using dictation?

**Imagining Perfect Speech Recognition**

For the following questions, imagine that in the future, digital devices can understand your speech perfectly.

Q: How interested would you be then in using speech input in general to control your digital devices?Why?

- Extremely interested
- Very interested
- Somewhat interested
- Not very interested
- Not at all interested

Q: How interested would you be then in using voice assistants? Why?

- Extremely interested
- Very interested
- Somewhat interested
- Not very interested
- Not at all interested

Q: How interested would you be then in using voice dictation as another option for text entry compared to keyboard input? Why?

- Extremely interested
- Very interested
- Somewhat interested
- Not very interested
- Not at all interested

Q: Are there any other ways that you use speech to control your digital devices? *If so, please describe.*

Q: Which speech-based features are they most interested in or feel would be most beneficial to you?



\end{document}